\documentclass{aa}  

\usepackage{graphicx}
\usepackage{subcaption}
\usepackage[T1]{fontenc}
\usepackage{graphicx}
\usepackage{multirow}
\usepackage{xcolor}
\usepackage{color}
\usepackage{amsmath}    
\usepackage{txfonts}
\usepackage{multirow}
\usepackage{notes2bib}
\usepackage{stfloats}
\usepackage{float}
\usepackage{txfonts}
\usepackage{hyperref}
\usepackage{comment}
\usepackage{wrapfig}
\usepackage{lscape}
\usepackage{rotating}
\usepackage{soul}
\def\INSPIRE{\mbox{{\tt INSPIRE}}}
\newcommand{\Reff}{$\mathrm{R}_{\mathrm{e}\,}$}
\newcommand{\Mstar}{M$_{\star}\,$}

\newcommand{\kms}{km/s}%$^{-1}$}
\newcommand{\Msun}{M$_{\odot}\,$}

\usepackage[normalem]{ulem}
\newcommand{\ppxf}{{\tt pPXF}}

\definecolor{darkgreen}{rgb}{0.09, 0.45, 0.27}
\definecolor{amber(sae/ece)}{rgb}{1.0, 0.49, 0.0}
\defcitealias{Tortora+18_UCMGs}{T18}
\defcitealias{Scognamiglio20}{S20}
\defcitealias{Ferre-Mateu+17}{F17}
\defcitealias{Spiniello20_Pilot}{\INSPIRE\, Pilot}
\defcitealias{Spiniello+21}{\INSPIRE\, DR1}

\begin{document} 

%%%%%  TITLE %%%%%
   \title{INSPIRE: INvestigating Stellar Population In RElics III. Second data release (DR2)\thanks{The data described in this paper are publicly available via the ESO Phase 3 
Archive Science (\url{https://archive.eso.org/scienceportal/home?data_collection=INSPIRE})}: testing the systematics on the stellar velocity dispersion}

%%%%% AUTHORS %%%%%
   \author{G.~D'Ago\inst{\ref{puc}}\fnmsep\thanks{Corresponding author: {\tt gdago@astro.puc.cl}} 
   %GROUP 1 
   \and C. Spiniello\inst{\ref{oxf},\ref{inaf_naples}}
   \and L.~Coccato\inst{\ref{eso}}
   \and C.~Tortora\inst{\ref{inaf_naples}}
\and F.~La Barbera\inst{\ref{inaf_naples}}
%   %GROUP 2
%      %GROUP 3 
   \and M.~Arnaboldi\inst{\ref{eso}}
   \and D.~Bevacqua\inst{\ref{insubria}}
   \and A.~Ferr\'e-Mateu\inst{\ref{iac},\ref{lalaguna}}
    \and A.~Gallazzi\inst{\ref{inaf_arcetri}}
    \and J.~Hartke\inst{\ref{eso2},\ref{oxf}}
    \and L.~K.~Hunt\inst{\ref{inaf_arcetri}}
  \and
  I.~Mart\'in-Navarro\inst{\ref{iac}}
  \and
N.R.~Napolitano\inst{\ref{china}}
  \and \\
  C.~Pulsoni\inst{\ref{mpe}}
   \and M.~Radovich\inst{\ref{inaf_padova}}
\and  
P.~Saracco\inst{\ref{brera}}
\and
D.~Scognamiglio\inst{\ref{bonn},\ref{jpl}} 
\and S.~Zibetti\inst{\ref{inaf_arcetri}}
}

   \institute{ 
   Instituto de Astrof\'{i}sica, Pontificia Universidad Cat\'olica de Chile, Av. Vicu\~na Mackenna 4860, 7820436 Macul, Santiago, Chile\label{puc}
    \and
    Sub-Dep. of Astrophysics, Dep. of Physics, University of Oxford, Denys Wilkinson Building, Keble Road, Oxford OX1 3RH, UK\label{oxf}
    \and
    INAF -  Osservatorio Astronomico di Capodimonte, Via Moiariello  16, 80131, Naples, Italy\label{inaf_naples}
    \and 
    European Southern Observatory,  Karl-Schwarzschild-Stra\ss{}e 2, 85748, Garching, Germany\label{eso}
    \and 
    Universita degli studi dell’Insubria, via Valleggio 11, 22100 Como, Italy\label{insubria}
    \and
    Instituto de Astrof\'isica de Canarias, V\'ia L\'actea s/n, E-38205 La Laguna, Tenerife, Spain\label{iac}
    \and
    Departamento de Astrof\'isica, Universidad de La Laguna, E-38205 La Laguna, Tenerife, \label{lalaguna}
    \and 
    European Southern Observatory, Alonso de Cordova 3107, Vitacura,
    Casilla 19001, Santiago de Chile, Chile\label{eso2}
    \and 
    INAF - Osservatorio Astronomico di Arcetri, Largo Enrico Fermi 5, 50125, Firenze, Italy\label{inaf_arcetri}  
    \and
    School for Physics and Astronomy, Sun Yat-sen University, Guangzhou 519082, Zhuhai Campus, China\label{china}
    \and
    Max-Planck-Institut f\"{u}r  extraterrestrische Physik, Giessenbachstrasse, 85748 Garching, Germany\label{mpe}
    \and     
    INAF - Osservatorio astronomico di Padova, Vicolo Osservatorio 5, I-35122 Padova, Italy\label{inaf_padova}
     \and
     INAF - Osservatorio Astronomico di Brera, via Brera 28, 20121 Milano, Italy\label{brera}
     \and
    Argelander-Institut f\"{u}r Astronomie, Auf dem H\"{u}gel 71, D-53121 - Bonn, Germany\label{bonn}
    \and
    Jet Propulsion Laboratory, California Institute of Technology, 4800 Oak Grove Drive, Pasadena, CA 91109, USA\label{jpl}
    }

   \date{Received November 23, 2022; accepted January 23, 2023}
   
   \titlerunning{INSPIRE DR2: testing the systematics on the stellar velocity dispersion}
   \authorrunning{G. D'Ago et al.}

  \abstract
   {The project called INvestigating Stellar Population In RElics (\INSPIRE) is based on VLT/X-Shooter data from the homonymous on-going ESO Large Program. It targets 52 ultra-compact massive galaxies at $0.1<z<0.5$ with the goal of constraining their kinematics and stellar population properties in great detail and of analysing their relic nature. }
   {This is the second \INSPIRE\ data release (DR2), comprising 21 new systems with observations completed before March 2022. For each system, we release four one-dimensional (1D) spectra to the ESO Science Archive, one spectrum for each arm of the X-Shooter spectrograph. They are at their original resolution. We also release a combined and smoothed spectrum with a full width at half maximum resolution of 2.51\AA. In this paper, we focus on the line-of-sight velocity distribution, measuring integrated stellar velocity dispersions from the spectra, and assessing their robustness and the associated uncertainties.}
   {For each of the 21 new systems, we systematically investigated the effect of the parameters and set-ups of the full spectral fitting on the stellar velocity dispersion ($\sigma$) measurements. In particular, we tested how $\sigma$ changes when several parameters of the fit as well as  the resolution and spectral coverage of the input spectra are varied.}
   {We found that the effect that causes the largest systematic uncertainties on $\sigma$ is the wavelength range used for the fit, especially for spectra with a lower signal-to-noise ratio (S/N $\leq 30$). When using blue wavelengths (UVB arm) one generally underestimates the velocity dispersion (by $\sim15$ \kms). The values obtained from the near-IR (NIR) arm present a larger scatter because the quality of the spectra is lower. We finally compared our results with those in literature, finding a very good agreement overall.} 
   {Joining results obtained in DR1 with those presented here, \INSPIRE\ contains 40 ultra-compact massive galaxies, corresponding to  75\% of the whole survey. 
   By plotting these systems in a stellar mass--velocity dispersion diagram, we identify at least four highly reliable relic candidates among the new systems. Their velocity dispersion is larger than that of normal-sized galaxies of similar stellar mass.}


   \maketitle
%%%%%%%%%%%%%%%%% BODY OF PAPER %%%%%%%%%%%%%%%%%%

\section{Introduction}
The most massive and oldest galaxies in the Universe (early-type galaxies, ETGs) play a fundamental role in the process of structure formation because they account for more than half of the total mass in the Universe \citep{Blumenthal+84_nature}. However, the details of their formation and evolution are a contentious question in present-day extragalactic astrophysics and cosmology. Recently, a two-phase formation scenario has been proposed to explain their mass assembly \citep{Naab+09, Oser+10, Oser+12, Hilz+13, Rodriguez-Gomez+16}. A few billion years after the Big Bang, intense and fast star formation episodes ($\tau\sim100$ Myr, star formation rate $\ge1000$ $M_{\odot}\, yr^{-1} $) created ultra-compact and massive objects. 
When these galaxies are observed to be passive, they are usually referred to as red nuggets \citep{Damjanov+11}. Subsequently, a second longer phase, dominated by accretion, mergers, and gas inflows, drove structural evolution and size growth, shaping the local giant elliptical galaxies we observe today \citep{Daddi+05, vanDokkum+08,Buitrago+18_compacts}. 
The stars that formed during the first phase become then the core of today’s ETGs, occupying their innermost regions \citep{Ferre-Mateu+19,Pulsoni21, Barbosa21}. 
A fraction of these red nuggets could also end up in the bulges of massive nearby spiral galaxies \citep{delaRosa16, Costantin21}. 

However, the exact formation mechanism of red nuggets and the physics that transform them into today’s massive galaxies is far from clear, and these questions have a significance well beyond the context of the size evolution of quiescent galaxies. The dense centres of local massive galaxies host the most massive black holes in the Universe \citep{Kormendy+13,Ferre-Mateu+15}. Their chemical enrichment history is extremely different from that of the Milky Way \citep{Worthey+94}, and their stars may have formed with a bottom-heavy (i.e. dwarf-rich) stellar initial mass function (IMF; \citealt{Conroy13, Martin-Navarro+15_IMF_variation, Sarzi+18,Barbosa21}).

In local ETGs, the material accreted in the second phase unfortunately contaminates the in situ (i.e. first phase) component that encodes the information about high-redshift baryonic processes, affecting its spatial and orbital distributions. In turn, obtaining spectra of high-$z$ red nuggets with a signal-to-noise ratio (S/N) that is high enough to perform detailed stellar population analyses would require prohibitive integration times with current instrumentation. 
Because of the stochastic nature of mergers, a small fraction of red nuggets fortunately survived without experiencing any interactions. They are now massive compact relic galaxies \citep{Trujillo+09_superdense}. Relics are thus the only objects that allow us to study the physical processes that shaped the mass assembly of galaxies in the high-$z$ universe with the same amount of detail achievable in the nearby Universe.

So far, only three relics have been found and were fully characterised in the local Universe: NGC1277 \citep{Trujillo+14, Martin-Navarro+15_IMF_relic, Beasley+18}, Mrk 1216, and PGC 032873 \citep{Ferre-Mateu+17}. These three objects all show a very quick (time-scales <1 Gyr) high-$z$ star formation and thus are consistent with having stars with a very old mean mass-weighted age ($\sim13$ Gyr, almost as old as the Universe). Moreover, their morphology, kinematics, and density profiles perfectly resemble those of $z > 2$ red nuggets \citep{Ferre-Mateu+17}. 

With the effort of enlarging the number of confirmed relics and also detecting them at higher redshifts, we started the project called INvestigating Stellar Population In RElics (\INSPIRE), which aims at building the first large catalogue of relics at $0.1<z<0.5$ \citep{Spiniello20_Pilot, Spiniello+21}.  
The (relatively $\sim30$) high S/N and wide wavelength (from UVB to NIR) spectra from the X-Shooter spectrograph (XSH; \citealt{Vernet11})  now allow us to infer the stellar kinematics and population properties (age, metallicity, elemental abundance, and low-mass end of the IMF slope) of a sample of ultra-compact massive galaxies (UCMGs) at $0.1<z<0.5$ with stellar masses $M_{\star}>6 \times 10^{10} M_{\odot}$ and effective radii \Reff<2 kpc. 

This is the second of the three planned yearly data releases, in which we analyse the spectra of 21 systems whose observations were completed before March 2022\footnote{The data are publicly available through the ESO Science Archive, \url{https://archive.eso.org/scienceportal/home?data_collection=INSPIRE}.}.  
Here, we focus on analysing the line-of-sight velocity distribution (LOSVD) of the 21 new systems, and, in particular, on measuring the integrated stellar velocity distribution ($\sigma$).  The latter is often used as proxy for the total mass of the galaxy, and it has been used in the literature to select relic candidates \citep[e.g.][]{Saulder+15_compacts}. 
Moreover, in DR1, we found that relics and especially extreme relics have larger integrated $\sigma$ than non-relics and normal-sized galaxies of similar stellar masses. Hence, we need to ensure that based on the quality and wavelength range of our spectra, we can robustly measure the integrated stellar velocity dispersion values from them. 
After demonstrating that the $\sigma$ can be securely inferred from medium-S/N spectra, this can be used as a selection criterion, together with the \Reff\ and stellar mass, to pre-select good relic candidates from on-going and future wide-sky surveys such as the Galaxy Evolution Survey with the 4-metre Multi-Object Spectrograph Telescope \citep[4MOST; ][]{4MOST_2019}. This paper therefore presents a systematic and quantitative analysis of all the parameters and code set-ups that might bias and influence the stellar velocity dispersion measurements. In addition, it presents the first effort to show that the stellar velocity dispersion is not higher for all compact objects, as was proposed by \citet{Saulder+15_compacts}, for example, but only for a sub-sample of them. These probably are the most reliable relic candidates. 

The paper is organised as follows. The sample and current status of the observations, as well as previous results obtained with \INSPIRE,\ are described in Sect.~\ref{sec:data}. The data reduction and analysis, including the extraction of the 1D spectra, the telluric correction on visual (VIS) and NIR arms, and the combination of the three arms, are described in Sect.~\ref{sec:data_analysis}, and the kinematical analysis and results are presented in Sect.~\ref{sec:losvd_tests}. In Sect.~\ref{sec:meas_comparison} we compare velocity dispersion values obtained from X-Shooter spectra to those inferred from GAMA\footnote{the Galaxy And Mass Assembly survey} spectra for ten objects in common. 
In Sect.~\ref{sec:mass-sigma} we present the stellar mass-velocity dispersion plot for DR1 and the DR2 \INSPIRE\ objects. 
Finally, we present our conclusions and outline the future development of \INSPIRE\ in  Sect.~\ref{sec:conclusions}. 

Throughout the paper, we assume a standard $\Lambda$-cold dark matter ($\Lambda$CDM)  cosmology with $H_0=69.6$ \kms\ Mpc$^{-1}$, $\Omega_{\mathrm{vac}} = 0.714,$ and $\Omega_{\mathrm{M}} = 0.286$ \citep{Bennett14}.

\section{INSPIRE project status and sample}
\label{sec:data}
\begin{figure}
\includegraphics[width=9cm]{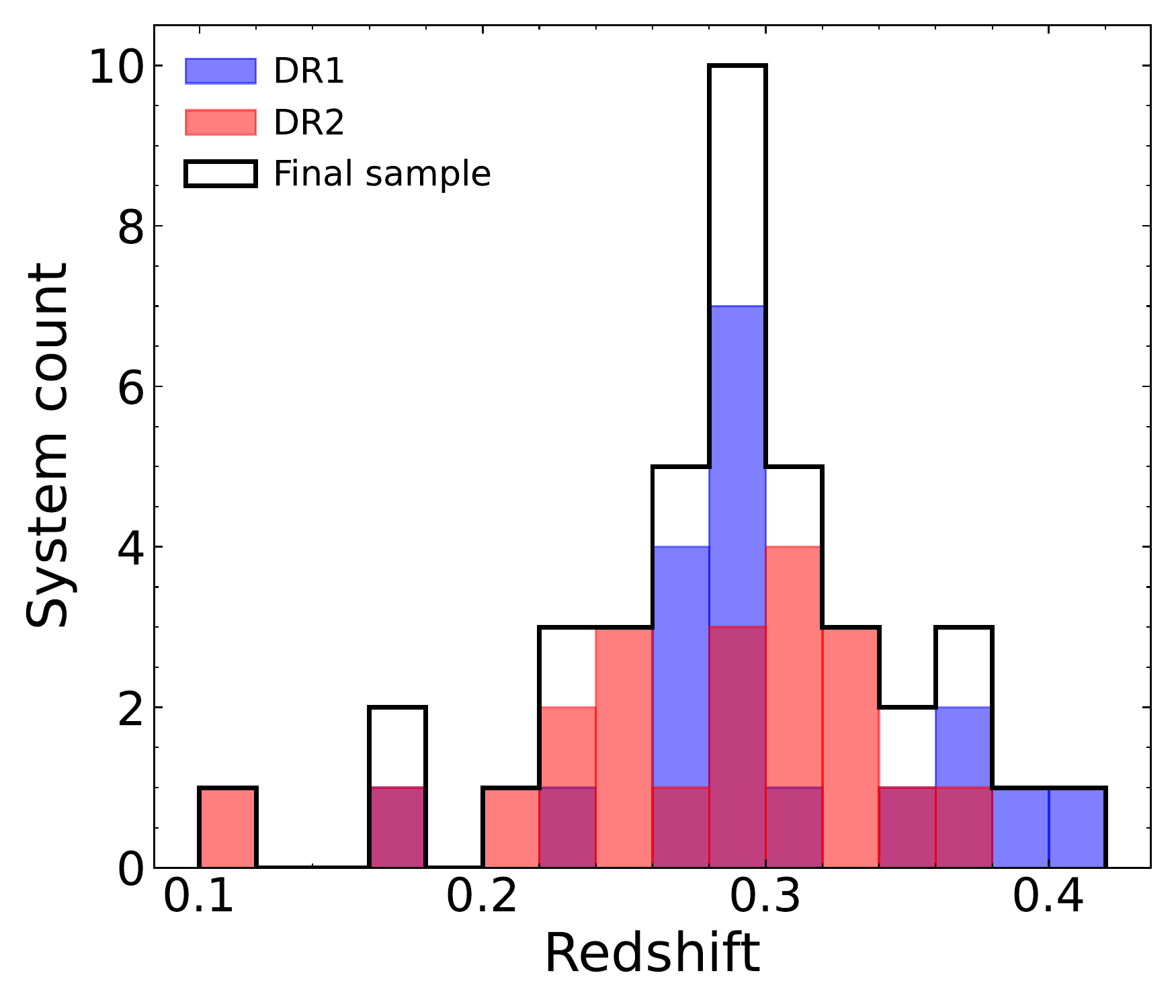}
\caption{Redshift distributions for the \INSPIRE\, targets in DR1 (blue), DR2 (red), and for the final sample (grey).}
\label{fig:obj_distrib}
\end{figure}

\INSPIRE\, is based on an on-going ESO Large Program (LP, ID: 1104.B-0370, PI: C. Spiniello) that started in P104 (October 2019) with the aim to spectroscopically follow up 52 UCMGs at  redshift $0.1<z<0.5$ with the XSH spectrograph \citep{Vernet11}.
The sample was collected through a dedicated observational effort to find and spectroscopically confirm as many UCMGs in the Kilo Degree Survey \citep[KiDS; ][]{Kuijken11} as possible \citep{deJong+15_KiDS_paperI, Kuijken19_KIDSDR4}. The results of this census for UCMGs were presented in \citet{Tortora+16_compacts_KiDS, Tortora+18_UCMGs}, and \citet{Scognamiglio20}, hereafter T16, T18, and S20. 
These objects are the perfect candidates to host very old stars as their $g-i$ broad-band colours are compatible with the colour of a stellar population with an integrated age $\ge8$ Gyrs (considering a solar, super-solar, and sub-solar metallicity; see Fig.~1 in \citetalias{Spiniello+21}). They also all have remarkably small sizes (\Reff$<2$ kpc) and high stellar masses (\Mstar$>6\times10^{10}$\Msun).  The redshift window covered by the sample, shown in Fig.~\ref{fig:obj_distrib}, is $0.1<z<0.5,$ with a peak at $z\sim0.28$. 

At the time of writing, 148 of the total 154 hours have been delivered by ESO. Forty systems were completely observed by the end of the ESO Period 108 (31 March 2022). Of these, 19 have been publicly released as part of the  \citetalias{Spiniello+21}, whereas the remaining 21 constitute the data release presented here that is publicly released via the ESO Phase 3 Science Archive\footnote{https://doi.org/10.18727/archive/36}.

We briefly describe the survey and observation strategy below. A more detailed description can be found in DR1. 
Depending on the $r$-band surface brightness luminosity ($\langle\mu_{\mathrm {e}}\rangle$) and aperture magnitudes ($\text{mag}_r$), both taken from the KiDS survey DR4 catalogue \citep{Kuijken19_KIDSDR4}, the exposure times on target varied from 2810 to 11240 seconds. This allowed us to obtain an integrated 1D spectrum with an S/N that was good enough (S/N $\ge15$ per \AA \,) to constrain the stellar age, metallicity, and [Mg/Fe] abundance of the stellar populations and hence confirm the relic nature of the UCMGs. 
For each of the targets, structural parameters were computed from KiDS images in $g,r,i$ bands and stellar masses inferred from spectral energy distribution (SED) fitting in $ugri$ bands. These numbers were reported in \citetalias{Tortora+18_UCMGs} and \citetalias{Scognamiglio20}. 

The slit widths that we chose for the UVB, VIS, and NIR arms are $1\arcsec.6$, $1\arcsec.5,$ and $1\arcsec.5$ respectively. The position angles (P.A.) of the slit were always oriented along the major axis of the galaxies, taken from \citetalias{Tortora+18_UCMGs} or \citetalias{Scognamiglio20}. The observations were carried out in nodding mode, with a dithering scheme consisting of multiple frames shifted by a small amount from the slit centre to facilitate a proper sky subtraction. Similarly to DR1, the seeing during the observations ranged between $0\arcsec.85$ to $1\arcsec.2$, with a median value of $\sim1\arcsec$. 

The feasibility of the method and techniques were extensively tested in \citet[][hereafter \INSPIRE\, Pilot]{Spiniello20_Pilot}, where the first three objects of the survey were presented. 
Then, in \citet[][hereafter \INSPIRE\, DR1]{Spiniello+21}, we applied the same routines to 19 systems that were fully observed until March 2020. 
We confirmed ten new relic galaxies, demonstrating that they had formed more than 75\% of their stellar mass at $z>2$, hence extending the number by a factor of 3.3 and pushing the redshift up to $z\sim0.5$. The remaining nine systems showed a longer star formation. 

An important result initially proposed by \citet{Ferre-Mateu+17} that was confirmed in DR1 is the degree of relicness: Some of the relics were already fully assembled in terms of stellar mass soon after the Big Bang (BB) and before the end of the first formation phase ($z\sim2$, extreme relics), whereas other relics formed a (high) fraction of their stars through a starburst in a very short time at high $z$ but then had subsequent lower-z star formation events (relics).  
Hence, the star formation history (SFH) of relics can be more or less extreme, and this might correlate with other morphological and stellar characteristics and possibly with the environment in which they live. 

As part of this DR2, we release the 1D NIR spectra of the 19 galaxies that were presented in DR1  to the ESO Phase 3 Science Archive, together with 1D UVB-VIS-NIR spectra of the 21 new systems. In particular, we obtain and release three fully reduced and flux-calibrated 1D extracted spectra per galaxy, one for each arm of the detector (UVB, VIS, and NIR), at their original resolutions. In addition, we combine the arms, after smoothing everything to a common resolution with a full width at half maximum  (FWHM) of $\sim2.51\AA$ and release the final spectrum as an additional data product. 

The \INSPIRE\, DR2 targets along with their coordinates 
are listed in Table~\ref{tab:sample}. We list the final exposure time and the P.A. of the slit, as well as photometric properties ($r$-band magnitudes and surface brightness), \Reff\ derived as the median of the quantities obtained from $g$, $r$ and $i$-band KiDS images and stellar masses, from SED fitting in $ugri$ bands. 
Finally, in the last column of the table, we report the sample from which the object was taken. 
Ten of the systems analysed in \citetalias{Tortora+18_UCMGs} or \citetalias{Scognamiglio20} were also independently observed with the AAOmega spectrograph and are part of the GAMA second data release database (DR2; \citealt{GAMA_DR2_Liske15}) or fourth data release (DR4;  \citealt{Gama_DR4_Driver22}). For these ten galaxies, we can directly compare the kinematic results obtained from the GAMA spectra with those computed from our spectra with their better resolution and higher S/N XSH (see Sec.~\ref{sec:meas_comparison}). 

\begin{table*}
\centering
\begin{tabular}{ccccccccccr}
\hline\hline
  \multicolumn{1}{c}{ID} &
  \multicolumn{1}{c}{RA} &
  \multicolumn{1}{c}{DEC} &
  \multicolumn{1}{c}{Exp.T.} &
  \multicolumn{1}{c}{P.A.} &
  \multicolumn{1}{c}{mag$_{r}$} &
  \multicolumn{1}{c}{$\langle\mu_{\mathrm {e}}\rangle_r$}&
 \multicolumn{1}{c}{$\langle\mathrm{R}_{\mathrm{e}}\rangle\,$} &
  \multicolumn{1}{c}{$\langle\mathrm{R}_{\mathrm{e}}\rangle\,$} &
  \multicolumn{1}{c}{M$_{\star}$} &
  \multicolumn{1}{c}{SAMPLE} \\
   
  \multicolumn{1}{c}{KiDS } &
  \multicolumn{1}{c}{(deg)} &
  \multicolumn{1}{c}{(deg)} &   
  \multicolumn{1}{c}{(sec)} &
  \multicolumn{1}{c}{(deg) } &
  \multicolumn{1}{c}{(AB)} &
  \multicolumn{1}{c}{(AB)} &
  \multicolumn{1}{c}{($\arcsec$) } &
  \multicolumn{1}{c}{(kpc)} &
  \multicolumn{1}{c}{($10^{11}$M$_{\odot}$)}  &
  \multicolumn{1}{c}{ } \\
\hline

J0844+0148 & 131.0553886 & +1.8132204  &  11240 & $-$37.4  & 19.78  & 18.53 & 0.26 & 1.14 & 0.71 & S20/GAMA\\ 
J0904-0018 & 136.0518949 & -0.3054848  &  5620  & $-$96.6  & 19.11  & 18.06 & 0.26 & 1.16 & 1.3  & S20/GAMA\\ 
J0909+0147 & 137.3989150 & +1.7880025  &  5620  & 10.3   & 18.68  & 16.05 & 0.30 & 1.05 & 1.05 & T18/GAMA\\
J0917-0123 & 139.2701850 & $-$1.3887918  &  11240 & $-$62.9  & 19.21  & 17.99 & 0.27 & 1.37 & 2.19 & S20\\ 
J0920+0126 & 140.1291393 & +1.4431610  &  11240 & $-$115.6 & 19.52  & 18.82 & 0.33 & 1.51 & 0.98 & S20/GAMA\\
J1026+0033 & 156.7231818 & +0.5580979  &  5620  & $-$83.1  & 17.39  & 16.98 & 0.34 & 1.02 & 1.48 & SDSS\\ 
J1040+0056 & 160.2152308 & +0.9407580  &  11240 & 53.4   & 19.52  & 18.85 & 0.31 & 1.29 & 0.93 & S20\\ 
J1114+0039 & 168.6994335 & +0.6510299  &  8430  & 34.0   & 19.0   & 17.89 & 0.34 & 1.52 & 1.62 & S20\\ 
J1128-0153 & 172.0885023 & $-$1.8890642  &  8430  & $-$2.9   & 18.56  & 17.94 & 0.35 & 1.27 & 1.30  & T18\\ 
J1142+0012 & 175.7023296 & +0.2043419  &  2810  & $-$84.8  & 17.02  & 17.90 & 0.71 & 1.40 & 0.84 & S20/GAMA\\
J1154-0016 & 178.6922828 & $-$0.2779248  &  8430  & 25.4   & 19.52  & 18.28 & 0.22 & 1.06 & 0.64 & T18/GAMA\\ 
J1156-0023 & 179.2186145 & $-$0.3946597  &  5620  & 15.8   & 18.83  & 17.01 & 0.26 & 1.04 & 1.39 & T18/GAMA\\
J1202+0251 & 180.5132277 & +2.8515452  &  8430  & $-$70.6  & 19.43  & 18.53 & 0.31 & 1.49 & 0.68 & S20\\ 
J1218+0232 & 184.7355807 & +2.5449139  &  5620  & 1.8    & 19.23  & 18.72 & 0.31 & 1.40 & 0.93 & S20\\ 
J1228-0153 & 187.0640987 & $-$1.8989049  &  5620  & $-$74.1  & 18.85  & 18.57 & 0.36 & 1.61 & 1.15 & S20\\ 
J1411+0233 & 212.8336012 & +2.5618381  &  8430  & 37.3   & 18.86  & 17.44 & 0.21 & 1.07 & 1.55 & S20/GAMA\\ 
J1436+0007 & 219.0481314 & +0.1217459  &  5620  & $-$100.6 & 18.27  & 18.27 & 0.39 & 1.40 & 1.15 & S20/GAMA\\ 
J2202-3101 & 330.5472803 & $-$31.0183808 &  8430  & $-$87.6  & 19.43  & 18.77 & 0.31 & 1.45 & 1.10 & T18\\ 
J2204-3112 & 331.2228147 & $-$31.2002605 &  8430  & $-$86.1  & 19.32  & 18.74 & 0.35 & 1.39 & 0.90 & T18 \\
J2257-3306 & 344.3966471 & $-$33.1144449 &  8430  & 3.5    & 19.42  & 17.09 & 0.29 & 1.18 & 0.93 & T18/GAMA\\ 
J2356-3332 & 359.1261248 & $-$33.5334748 &  11240 & $-$46.5  & 19.81  & 18.37 & 0.22 & 1.06 & 0.98 & T18\\ 
\hline
\hline
\end{tabular}
\caption{  \INSPIRE\, DR2 sample. We list from left to right the galaxy ID and coordinates, the exposure times and the position angles (along the major axis of the galaxy) of the XSH observations, the aperture magnitudes (MAG\_AUTO from the KiDS DR3 catalogue, corrected for extinction), the surface brightness luminosity averaged within the \Reff, both in $r$-band, \Reff\ in arcseconds and kiloparsec, computed as median of the quantities obtained from the $g,r,i$ bands,  and the stellar masses from the SED fitting. Finally, in the last column, we list the sample from which each object was taken. The six objects with a double reference were selected from T18(or S20), but were then also found in the GAMA DR4. }
\label{tab:sample}
\end{table*}

\section{Data reduction and analysis}
\label{sec:data_analysis}
\subsection{Data reduction and 1D extraction}
As already explained in the \citetalias{Spiniello+21}, we performed an ad hoc extraction of the one 1D spectra to take into account the fact that these galaxies are not spatially resolved and the spectra are dominated by seeing, as the \Reff\ of all objects in arcseconds (apparent sizes, on average \Reff$\sim0\arcsec.3$) are much smaller than the median seeing of the observations ($\sim 1\arcsec$ on average). 
Hence, we reduced the data using the ESO XSH pipeline (v3.5.3) under the ESO Reflex Workflow (\citealt{Freudling+13}, version 2.11.3), only up to the creation of the 2D spectral frames (one for each arm). Then, we subsequently used our own Python routines, developed for the  \citetalias{Spiniello20_Pilot} , which were used in  \citetalias{Spiniello+21}. We cannot use ESO internal data products because they only comprise the already extracted 1D spectra. 

Finally, for the VIS and NIR arms, we corrected all the spectra for telluric absorption lines using the code \textsc{molecfit} (\citealt{Smette15}, version 4.2), which was run with its interactive ESO Reflex workflow. 
The telluric correction was performed with the recipe \textit{molecfit\_model} that fits telluric absorption features on the telluric standard that is observed in the same night and with the same instrument set-up as the galaxies. After we determined the column densities of the various molecules in the spectrum, we constructed the telluric correction and took the difference in air-mass between the observations of the telluric standard and the galaxy into account. For this purpose, we used the recipe \textit{molecfit\_calctrans}. 

In previous papers of the \INSPIRE\, series, we have extracted spectra with two different approaches. On the one hand, we collapsed the whole slit, but weighted the pixels by their flux (following the optimal extraction approach described in \citealt{Naylor+98}).
Alternatively, as a  second approach, we also extracted the spectra of each galaxy from an aperture that contained more or less the same fraction of light for the different objects (R50, containing $\sim50$\% of the total light, but a mix from inside and outside the real \Reff, given the spatial resolution of the data; see \citetalias{Spiniello+21} for more details). The R50 approach is best when the \INSPIRE\, sample is to be compared with other galaxy samples from the literature because this is the most comparable aperture, at least in terms of light fraction, to that extracted at one \Reff\ for normal-size galaxies. In
\citetalias{Spiniello+21},  we proved that the extraction method does not change the kinematics and stellar population results, and hence it does not play a role in the relic confirmation. Therefore, in this DR2, we extracted spectra following the R50 approach alone.

\subsection{Arm combination and smoothing}
\label{sec:smoothing}
When deriving the stellar population parameters, it is important to use a wide wavelength range that allows breaking the age-metallicity degeneracy \citep{Worthey+94} and thus properly inferring the stellar population parameters. Hence, the three arms in the \INSPIRE\ data set must be combined. %the three arms must be combined. 
In order to do this, they must first be brought to the same final resolution. 
We point out that the UVB and VIS spectra have the same spatial and spectral sampling (scale = $0\arcsec.16/\text{px}$, $\delta\lambda\sim0.156$ \AA), while the NIR has lower resolution (scale = $0\arcsec.25/\text{px}$, $\delta\lambda\sim0.467$ \AA). 

First, we computed the redshift of each galaxy and independently identified the most prominent stellar absorption lines in the different arms.
The redshift values are always consistent between the arms, and in 20 out of 21 cases, they are also consistent with the values reported in \citetalias{Tortora+18_UCMGs} and \citetalias{Scognamiglio20} (within 0.0005, the nominal uncertainties on the redshifts). 
Only for J1218+0232 do we find a slightly higher redshift from the XSH spectrum with a higher S/N than that used by T18 and S20 ($\Delta z = z_{\text{XSH}} - z_{\text{S20}} = 0.0352$).

Then, we reported all the spectra for all the systems and in all arms to the same resolution at fixed FHWM and the same binning. We chose an FWHM$_{\mathrm{fin}} = 2.51$\AA, which is\, equal to that of the MILES single stellar population (SSP) models \citep{Vazdekis15} that we used for the kinematic analysis. 
For the smoothing, we adopted the same spectral convolution procedure as was successfully employed in the  \citetalias{Spiniello20_Pilot} and in the \citetalias{Spiniello+21}, based on the use of a Gaussian function with a variable sigma (following the prescription of \citealt{Cappellari17}). 

After smoothing and binning, we finally joined the three arms using data from the UVB up to $\lambda=5560$\AA\ (observed wavelength), from VIS up to $\lambda=10100$\AA,\ and from the NIR for redder wavelengths. The arms extend slightly further (UVB 5595\AA\ and VIS 10240 \AA) and overlap by  $\sim300\AA$. Anyway, we point out that the cut we set for each of the three arms was chosen in order to avoid extremely noisy wavelengths at the borders. This choice has no effect on the results presented in the paper because we measured the kinematics both from the single-arm spectra and from the combined ones.

The redshifts inferred from the combined and smoothed spectra are reported in the second column of Table~\ref{tab:spec_data}. 
They are fully consistent with those computed from the single-arms spectra. Finally, we brought all the spectra to $z=0$. 

\subsection{Signal-to-noise calculation}
To calculate the S/N of the 1D spectra, we followed the same recipe %already successfully 
as was used in DR1. We used the IDL code \texttt{}{DER\_SNR} \citep{Stoehr08}, which estimates the S/N directly from the flux, assuming that it is Gaussian distributed and uncorrelated in wavelength bins spaced two pixels apart.

We obtained three different estimates, one for each arm separately (across the entire wavelength, which slightly varies from one object to the next because their redshifts are different), and then, we also computed the arithmetic mean of the three.
These numbers are reported in Table~\ref{tab:spec_data}, along with the redshifts inferred from the final combined and smoothed spectra. We recall that the three independent estimates of the redshift obtained from the three single spectra are perfectly consistent with the estimate computed from the combined spectrum. 
For all the objects, the S/N increases from the UVB to the VIS, demonstrating the passive nature of the systems,  but then decreases again in the NIR; this is likely due to the noisier nature of the NIR spectra. 
We did not compute the S/N on the combined and smoothed spectra because the convolution causes a noise correlation across different pixels, and hence the assumption made by the \texttt{}{DER\_SNR} code is no longer valid. 

\section{Analysis of the line-of-sight velocity distribution}
\label{sec:losvd_tests}
To derive the LOSVD, and, in particular, to obtain the integrated stellar velocity  dispersion values, we used the Python-based penalised pixel-fitting software (\ppxf, v8.1.0;  \citealt{Cappellari04,Cappellari17}).
We only used an additive Legendre polynomial (ADEGREE) to correct for the continuum shape during the fit, but did not use a multiplicative one, nor did we use any regularisation (see \citealt{Cappellari17} for more details) because in this paper, we focus on the kinematics without computing stellar population parameters and obtaining the relic confirmation, which will be presented in a follow-up paper (Spiniello et al., in prep).

The main purpose of our analysis is to assess the robustness of the inferred velocity dispersion values that will be used in forthcoming INSPIRE papers as a proxy for the total mass in the dynamical modelling of UCMGs in general and relics in particular (Tortora et al., in prep). 
Moreover, the velocity dispersion also appears to be a good way to select high-confidence relic candidates because we found in the  \citetalias{Spiniello+21} that relics have a higher velocity dispersion than non-relics with similar sizes and stellar masses. 

In order to assess how solid the inferred $\sigma$ values are, we performed an extensive number of tests to determine the dependence of the kinematic measurements upon the input spectra and the various assumptions and parameters in the \ppxf\, fit: the ADEGREE and the moments of the LOSVD were left as free parameters, the wavelength range, and the masked pixels. We did not need to investigate on the effect of changing the stellar templates in the fitting procedure because this was already done in the \citetalias{Spiniello20_Pilot}. No detectable difference in the inferred $\sigma$ was found.
Therefore, we only fit with the EMILES models presented in  \citet{Vazdekis15}. We limited ourselves to the safe range of parameters, following the prescription of the authors of the models\footnote{The safe ranges are defined as follows: ages: $[0.063-17.8]$ Gyr, and metallicites: $[-2.32, 0.22]$, [$\alpha$/Fe]: $[0.0-0.4]$. See \url{http://research.iac.es/proyecto/miles/pages/ssp-models/safe-ranges.php} for more details.}. We used the models with a bimodal stellar IMF with a fixed slope of $\Gamma = 1.3$ and PADOVA00 theoretical isochrones \citep{Girardi00}. 
In a forthcoming publication of the \INSPIRE\, series, we will directly investigate the non-universality of the IMF \citep{Martin-Navarro-submitted}. In this analysis, we study the UVB+VIS+NIR and fit up to 10000 \AA\, (rest frame) alone, where IMF effects are weaker (since dwarf stars emit more strongly at redder wavelengths), and where the majority of the narrow, strong stellar absorption lines are. 
The three spectra separately in the three arms as well as the combined spectra are made public for the astronomical community and can be downloaded directly from the ESO Science Archive. They range from 2300\AA (2800) to 18000\AA\ (23000) for the highest (lowest) redshift in the sample.

\begin{table}
\centering
\begin{tabular}{l|c|cccc}
\hline\hline
  \multicolumn{1}{c|}{ID} &
  \multicolumn{1}{c|}{z} &
  \multicolumn{1}{c}{S/N}    &
  \multicolumn{1}{c}{S/N}    &
  \multicolumn{1}{c}{S/N}    &
  \multicolumn{1}{c}{S/N}    \\
  \multicolumn{1}{c|}{KiDS} &
  \multicolumn{1}{c|}{($\pm0.0005$) } &
  \multicolumn{1}{c}{UVB}    &
  \multicolumn{1}{c}{VIS}    &
  \multicolumn{1}{c}{NIR}    &
  \multicolumn{1}{c}{MEAN}    \\
\hline
J1142+0012 & 0.1077 & 57.9 & 124.1 & 69.8 & 83.9 \\
J1026+0033 & 0.1743 & 38.9 & 113.6 & 68.3 & 73.6 \\
J0909+0147 & 0.2151 & 20.7 & 75.3  & 45.1 & 47.0 \\
J1228-0153 & 0.2973 & 23.2 & 70.1  & 40.8 & 44.7 \\
J1128-0153 & 0.2217 & 21.1 & 69.2  & 37.0 & 42.4 \\
J1411+0233 & 0.3598 & 24.1 & 73.2  & 26.9 & 41.4 \\
J1436+0007 & 0.221  & 21.1 & 67.2  & 29.0 & 39.1 \\
J1156-0023 & 0.2552 & 22.6 & 60.9  & 31.8 & 38.4 \\
J0920+0126 & 0.3117 & 17.9 & 55.6  & 29.7 & 34.4 \\
J1114+0039 & 0.3004 & 19.5 & 54.0  & 28.4 & 34.0\\ 
J2204-3112 & 0.2581 & 14.4 & 54.1  & 24.5 & 31.0 \\
J0917-0123 & 0.3602 & 12.2 & 50.3  & 27.5 & 30.0 \\
J1040+0056 & 0.2716 & 11.5 & 46.7  & 31.4 & 29.9 \\
J1202+0251 & 0.3298 & 14.7 & 45.9  & 25.3 & 28.6 \\
J0844+0148 & 0.2837 & 12.9 & 45.0  & 28.0 & 28.6 \\
J0904-0018 & 0.2989 & 12.6 & 44.3  & 26.6 & 27.8 \\
J1154-0016 & 0.3356 & 16.6 & 42.8  & 23.5 & 27.6 \\
J2257-3306 & 0.2575 & 17.8 & 40.0  & 24.0 & 27.3 \\
J2202-3101 & 0.3185 & 13.1 & 47.6  & 20.5 & 27.1 \\
J1218+0232 & 0.308  & 14.6 & 42.0  & 22.7 & 26.4 \\
J2356-3332 & 0.3389 & 11.5 & 34.2  & 17.7 & 21.1 \\
\hline\hline
\end{tabular}
\caption{Spectroscopic properties of the \INSPIRE\ DR2 sample. We list from left to right the ID, the redshift computed from the combined spectra, the three estimates of the S/N (per \AA) from the three arms at their original resolution, and the arithmetic mean of these three values. }
\label{tab:spec_data}
\end{table}

\subsection{Changing the ADEGREE parameter}
\label{sec:adegree}
\begin{figure*}
    \centering
\includegraphics[width=18.5cm]{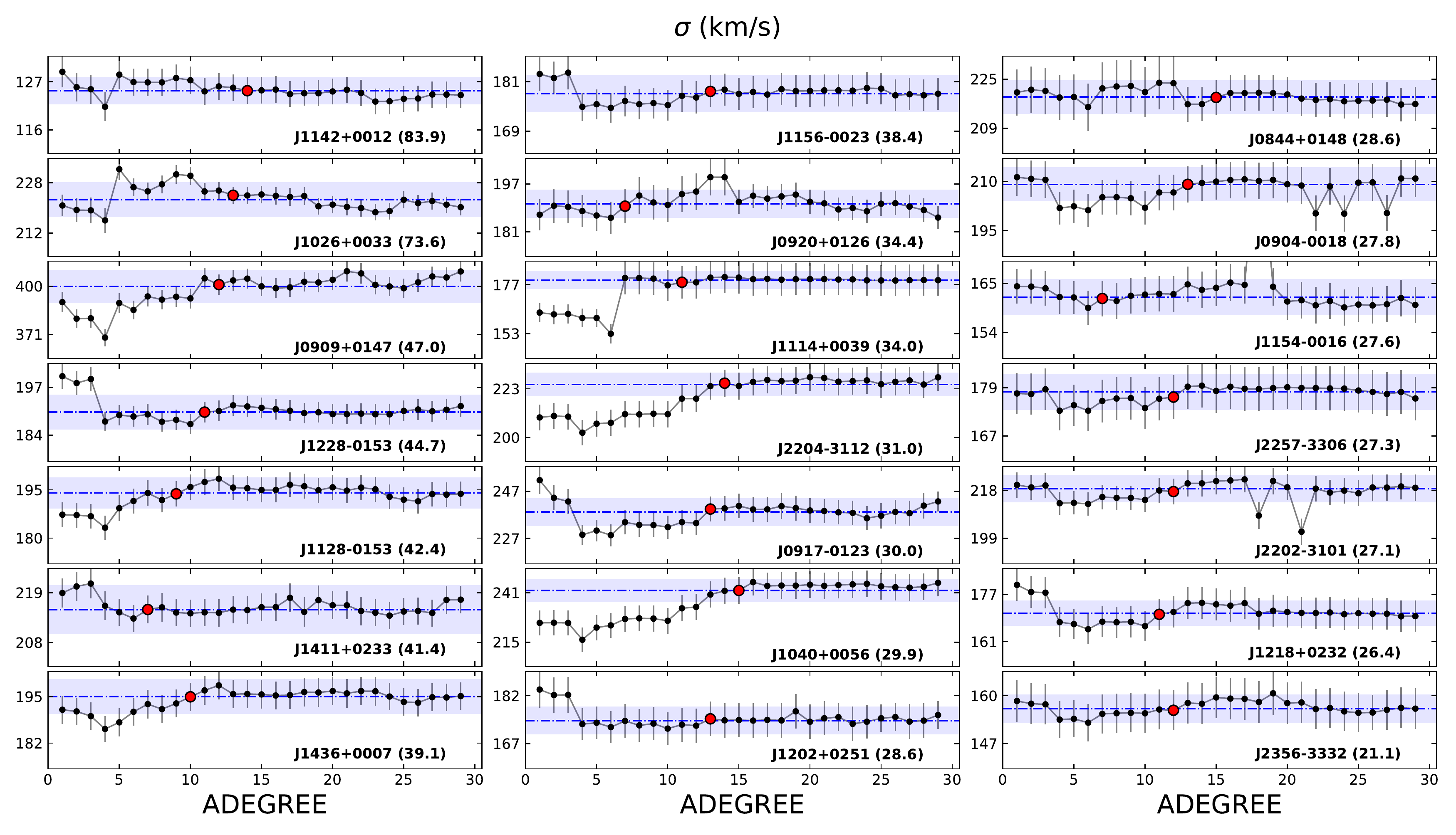}
\caption{ADEGREE test: Variation in the measured $\sigma$ as a function of the degree of the additive polynomial. The spectra are ordered from top to bottom and from left to right in descending S/N (shown in each panel in parentheses). The  larger red dots are the ADEGREE chosen for the fiducial fit, and the shaded blue region shows a variation on the $\sigma$ from the median value of $\pm2.5$\%.} 
\label{fig:adegree_test}
\end{figure*}

According to the \ppxf\, recommendations, when performing a full spectral fitting to obtain the LOSVD, only an additive Legendre polynomial should be used to correct for the continuum shape during the kinematic fit. The degree of the additive polynomial, which is regulated by the ADEGREE keyword in \ppxf,  might influence the final result on the integrated stellar velocity dispersion, especially in case of spectra with a low S/N. 
In previous \INSPIRE\ papers, we have tested a range of values for the ADEGREE parameter, always fixing it to the value that stabilises the results against changes of other parameters while at the same time minimizing the reduced $\chi'^{2}$ ($\chi^{2}$ divided by the number of good pixels used for the fit.). We decided to use the same ADEGREE value for all the systems (ADEGREE$\sim20$) to speed up the computational time, but risked to over-fit the noise affecting the stellar continuum in some cases, however. 
Here, we repeated this test in a more systematic way, using a broader range of values and more spectra. In particular, we tested ADEGREE values from 1 to 30 and found that at low polynomials degrees ($<5$), the inferred velocity dispersion is generally not stable. Then it reaches a plateau around a degree of 8-15, and in some cases, it then starts to wiggle again at very high degree values ($>17$). 
However, as expected, this plateau falls in a slightly different range of ADEGREE values for different systems, and this prevents us from choosing the same degree for all them.
In Fig.~\ref{fig:adegree_test} we indicate the fiducial choice of the ADEGREE value with a large red dot. This adopted choice for each system lies on the median $\sigma$ retrieved from the test, and it has two purposes: it represents a good central guess for the bootstrap routines we present in Sect.~\ref{sec:bootstrap} because it is far from the region in which $\sigma$ is strongly unstable. It also allows us to be sensitive to the uncertainties arising from the choice of a specific ADEGREE, hence assessing the error budget associated with this parameter. This is because the velocity dispersion values change slightly for degrees around the fiducial value.

Finally, we note that the S/N does not appear to depend on the input spectrum (shown in each panel), and the overall variation in $\sigma$ is generally about 5\%. This is shown in the figure as the shaded blue region.

\begin{figure*}
\centering
\includegraphics[width=18.5cm]{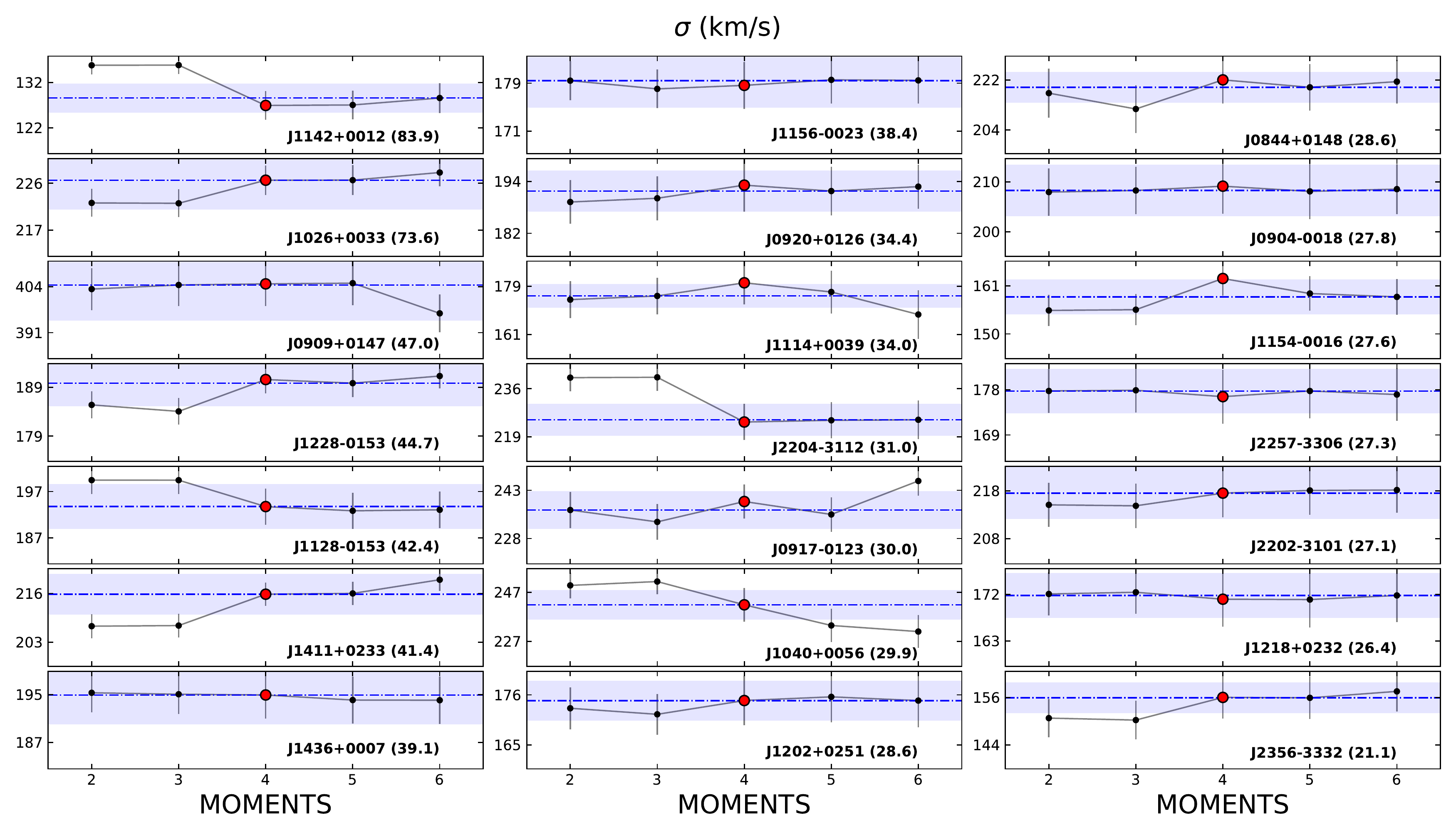}
\caption{MOMENTS test: Variation in the measured $\sigma$ as a function of the moments of the LOSVD that were constrained in the fit. The objects are in the same order as in Fig.~\ref{fig:adegree_test}, from high (top left) to low (bottom right) S/N.  In each panel, the galaxy ID and the S/N are plotted in the right bottom corner, the larger red dot is the MOMENT chosen for the fiducial fit (=4), and the shaded blue region shows a variation on the $\sigma$ from the median value of $\pm2.5$\%.}
\label{fig:moments_test}
\end{figure*}

\subsection{Changing the MOMENTS parameter}
\label{sec:moment}
The \ppxf\, allows us to constrain from two and up to six moments of the Gauss-Hermite parameterisation \citep{vanderMarel93}, using the keyword MOMENTS. In an ideal case, the higher moments ($h_3$,...,$h_6$) should be completely uncorrelated to the velocity (V) and velocity dispersion ($\sigma)$. However, when the LOSVD is not perfectly sampled by the data, performing the fit with a different number of moments might influence the resulting $\sigma$ values. 
Hence, we ran different fits changing the MOMENTS from two to six in order to test the effect of adopting different choices for the MOMENTS parameter. We show the resulting $\sigma$ values in Fig.~\ref{fig:moments_test}, where galaxies are again ordered according to the mean S/N. During this test, we used the fiducial value for the ADGREE parameter for each galaxy (i.e. the red dot in Fig.~\ref{fig:adegree_test}). No correlation with the S/N is found in this case either, and the $\sigma$ values inferred using different MOMENTS are very close to each other and in many cases are consistent within the errors. For the majority of systems, a small increase in $\sigma$ ($\sim5-10$ \kms) is observed from MOMENTS = 2 to MOMENTS = 4. Then, in 16 out of 21 cases the result is virtually unchanged when the number of fitted moments is further increased. In two (three) cases, $\sigma$ decreases (increases) when the number of moments in the fit is increased. 
Given the results of this test, we consider as fiducial values for the velocity dispersion the values computed with MOMENT=4 for all the 21 systems. In all but two cases, the variation with respect to the median value is always below 5\% (shaded region in the figure).

\begin{figure}
\centering
\includegraphics[width=9cm]{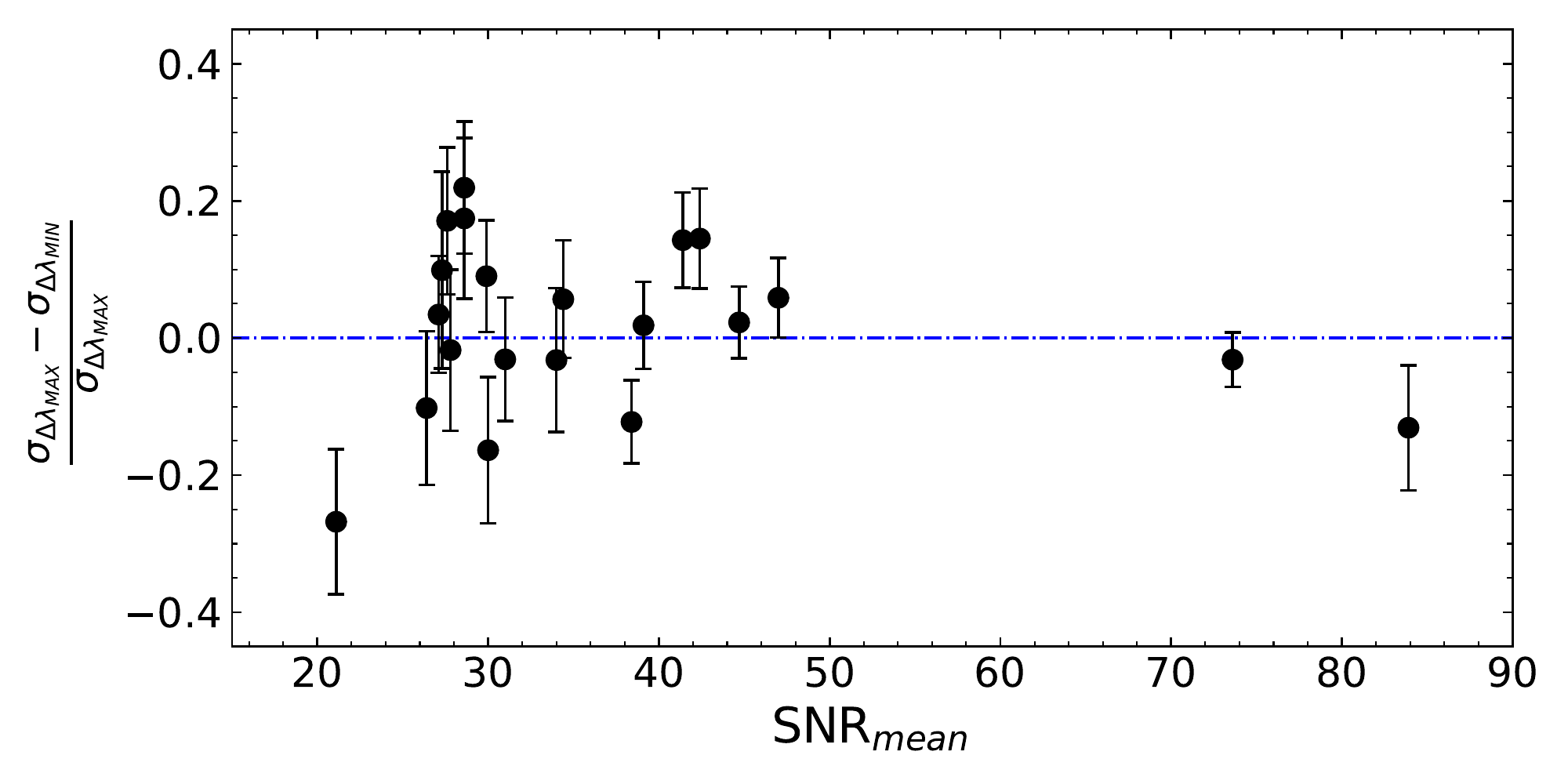}
\caption{Fit range test. The plot displays the relative difference in $\sigma$ between the measurements obtained by fitting the widest and narrowest wavelength range. For spectra with a lower S/N, the spread is larger and the variation in the velocity dispersion is in some cases as high as  20\%. }
\label{fig:fitrange_test}
\end{figure}

\subsection{Changing the wavelength range of the fitted region}
\label{sec:waverange}
Another effect we tested is the effect of the fitted spectral wavelength on the inferred $\sigma$ values. For this purpose, we repeated the fit many times with the fiducial ADEGREE and MOMENT parameters, but each time with a smaller wavelength range by systematically shifting the blue and the red ends limits of the fitted window. We started from the widest range, [3000$-$10000] \AA\ and decreased in steps of 200 \AA\ on each side of the fitting window down to the smallest range, [6000$-$7000] \AA.  
In this case, we find a dependence on the S/N: A larger variation in the inferred $\sigma$ values is observed in low-S/N spectra. This is visible from Fig.~\ref{fig:fitrange_test}, where we plot the relative difference in velocity dispersion between the value obtained by fitting the widest wavelength range and the value obtained fitting the narrowest range, against the S/N. 
For S/N$<35,$ the scatter is larger than 20\% ($\Delta\sigma/\sigma_{\text{max}}>0.2$), while it stabilises around that level or below it for spectra with higher S/N. 
For the  majority of the cases, higher values of $\sigma$ are found when the 4000 \AA\ break is masked out. This effect is smaller for high S/N spectra, however, where the results are much more stable against change in the fitted region. 

\begin{figure*}[htbp]
\centering
\includegraphics[width=18.5cm]{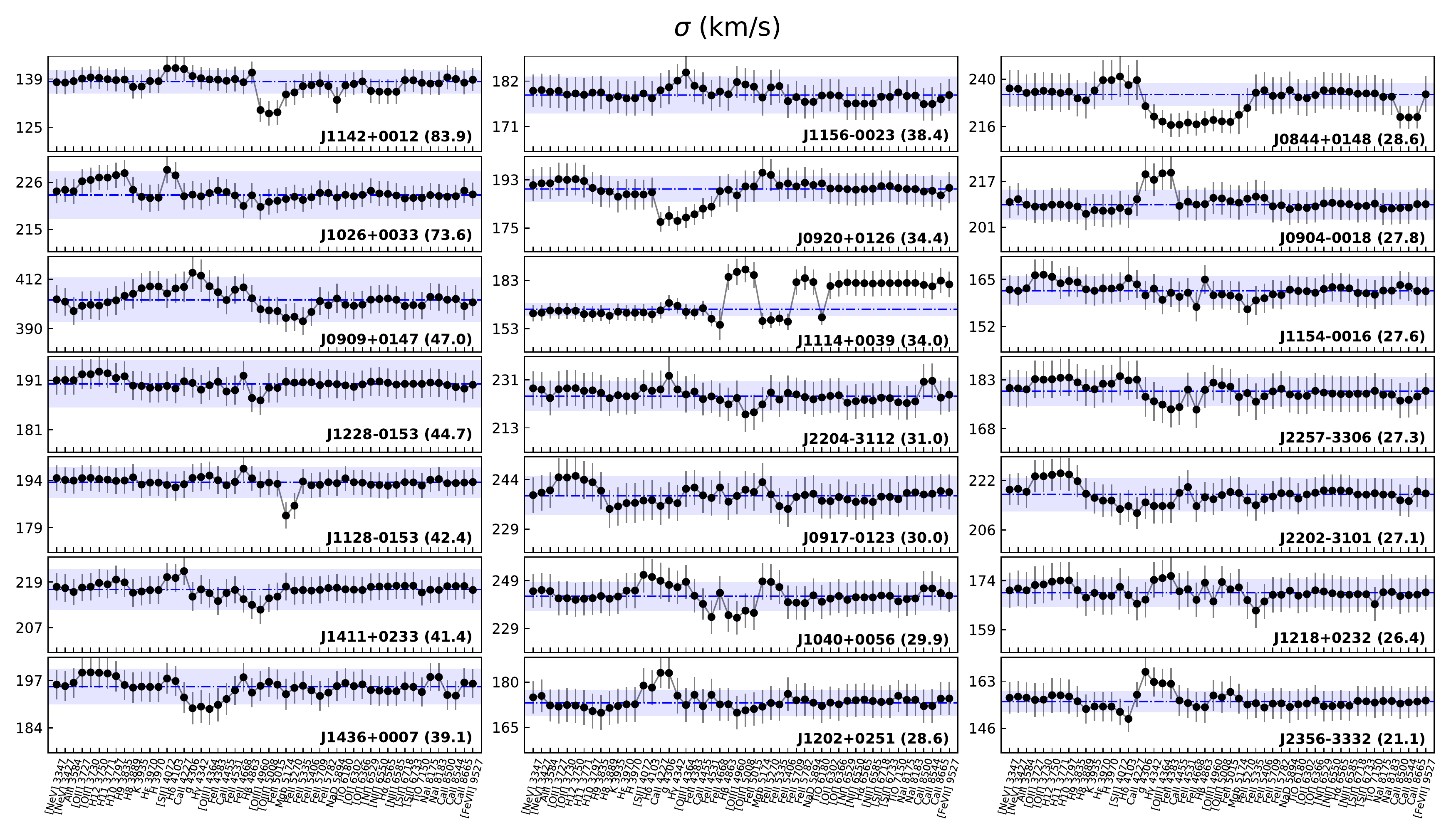}
\caption{Random masking test. The objects are in the same order as in Fig.~\ref{fig:adegree_test}, from high (top left) to low (bottom right) S/N. The figure shows the effect on the $\sigma$ of masking a window of 200 $\AA$
placed on each of the spectral lines falling within the fitting limits out from the \ppxf\ fit.}
\label{fig:fitmask_test}
\end{figure*}

To evaluate the effect of masking a certain line from the fit, we ran another test in which we repeated the fit on the same spectrum across the fiducial fitting window (3000$-$10000 \AA) 50 times, each time masking one strong  (absorption or emission) stellar  feature. 
We note that frequently, different lines are less than 200 \AA\ apart. We nevertheless ran the masking test one time for each line, always centreing the masking window on the corresponding line (sometimes partially masking nearby lines as well).  
The result of this test is shown in Fig.~\ref{fig:fitmask_test} where again we group the spectra according to their mean S/N. Within each group, the spectra were also ordered from the highest (top) to the lowest (bottom) S/N. In this case, a mild dependence on the S/N is found overall, although changes on a single-system level are also visible. 
In general, the two regions appear to produce a non-negligible effect on the $\sigma$ measurements for more than one galaxy: the region between 3800 and 4300 \AA,\ and that around the Mg$_b$ and Fe strong absorption line ($\sim5100$-$5400$ \AA). 
Overall, however, the differences in sigma ($\Delta\sigma$) are never larger than 10\%. 

\begin{table}
\begin{tabular}{l|ccccc}
\hline\hline
  \multicolumn{1}{c|}{ID} &
  \multicolumn{1}{c}{$\sigma$ (\kms)} &
  \multicolumn{4}{c}{$\Delta\sigma_{BOOT}$ (\kms)} \\

  \multicolumn{1}{c|}{KIDS} &
  \multicolumn{1}{c}{COMB}    &
  \multicolumn{1}{c}{$\Delta_{noise}$}   &
  \multicolumn{1}{c}{$\Delta_{adeg}$}    &
  \multicolumn{1}{c}{$\Delta_{mom}$} &
  \multicolumn{1}{c}{$\Delta_{wav}$}   \\
\hline
    J1142+0012 & 129 & $\pm$6 & $\pm$6 & $\pm$6 & $\pm$6\\ 
    J1026+0033 & 225 & $\pm$5 & $\pm$7 & $\pm$8 & $\pm$6\\ 
    J0909+0147 & 401 & $\pm$15 & $\pm$18 & $\pm$17 & $\pm$12\\ 
    J1228-0153 & 191 & $\pm$6 & $\pm$8 & $\pm$7 & $\pm$7\\ 
    J1128-0153 & 192 & $\pm$8 & $\pm$9 & $\pm$8 & $\pm$8\\ 
    J1411+0233 & 217 & $\pm$7 & $\pm$9 & $\pm$8 & $\pm$9\\ 
    J1436+0007 & 193 & $\pm$9 & $\pm$10 & $\pm$10 & $\pm$10\\ 
    J1156-0023 & 177 & $\pm$11 & $\pm$11 & $\pm$10 & $\pm$10\\ 
    J0920+0126 & 190 & $\pm$8 & $\pm$8 & $\pm$8 & $\pm$9\\ 
    J1114+0039 & 181 & $\pm$10 & $\pm$13 & $\pm$12 & $\pm$10\\ 
    J2204-3112 & 227 & $\pm$14 & $\pm$16 & $\pm$13 & $\pm$14\\ 
    J0917-0123 & 239 & $\pm$13 & $\pm$13 & $\pm$13 & $\pm$14\\ 
    J1040+0056 & 240 & $\pm$14 & $\pm$18 & $\pm$16 & $\pm$14\\ 
    J1202+0251 & 165 & $\pm$17 & $\pm$14 & $\pm$13 & $\pm$17\\ 
    J0844+0148 & 224 & $\pm$12 & $\pm$12 & $\pm$15 & $\pm$15\\ 
    J0904-0018 & 205 & $\pm$11 & $\pm$12 & $\pm$11 & $\pm$13\\ 
    J1154-0016 & 163 & $\pm$8 & $\pm$10 & $\pm$10 & $\pm$9\\ 
    J2257-3306 & 185 & $\pm$15 & $\pm$13 & $\pm$12 & $\pm$13\\ 
    J2202-3101 & 221 & $\pm$13 & $\pm$12 & $\pm$12 & $\pm$10\\ 
    J1218+0232 & 171 & $\pm$11 & $\pm$13 & $\pm$11 & $\pm$16\\ 
    J2356-3332 & 162 & $\pm$15 & $\pm$15 & $\pm$15 & $\pm$16\\ 

\hline\hline
\end{tabular}
\caption{Results of the kinematics bootstrap analysis for the \INSPIRE\ DR2 sample. We report from left to right the ID, the fiducial $\sigma$ resulting from the bootstrap analysis on the combined and smoothed spectra ($\sigma_{\text{COMB}}$), and the  uncertainties associated with the four bootstrap routines. Each subsequent column includes the randomisation of the previous test (see text for more details). }
\label{tab:bootstrap_data}
\end{table}

\subsection{Bootstrap}
\label{sec:bootstrap}
After assessing the effects of the main fit parameters individually, we set up different bootstrap routines, always repeating 250 \ppxf\ fits of the same spectrum, to quantitatively infer uncertainties on the velocity dispersion values and to combine all the tests we performed so far in a systematic way. Specifically, we started to randomise each time the flux of every single pixel according to a Gaussian distribution of the noise around the observed flux and computed the uncertainties associated with this ($\Delta_{noise}$). 
Then, keeping the randomisation of the noise, we also added a random selection of the ADEGREE within the range 1-30 in the bootstrap, and evaluated the uncertainties in this case as well ($\Delta_{adeg}$). 
Subsequently, we ran a third bootstrap in which we combined the noise randomisation and the ADEGREE randomisation, and we also randomly changed the moments of the LOSVD ($\Delta_{mom}$), always from two to six. 
We finally ran another bootstrap including the three effects described above and a randomisation of the fit limit ($\Delta_{wav}$), chosen within 2500 pixels from the blue and red end of the entire wavelength range. We followed the same approach as described in  Sec.~\ref{sec:waverange}. 

\begin{figure}
    \centering    \includegraphics[width=8cm]{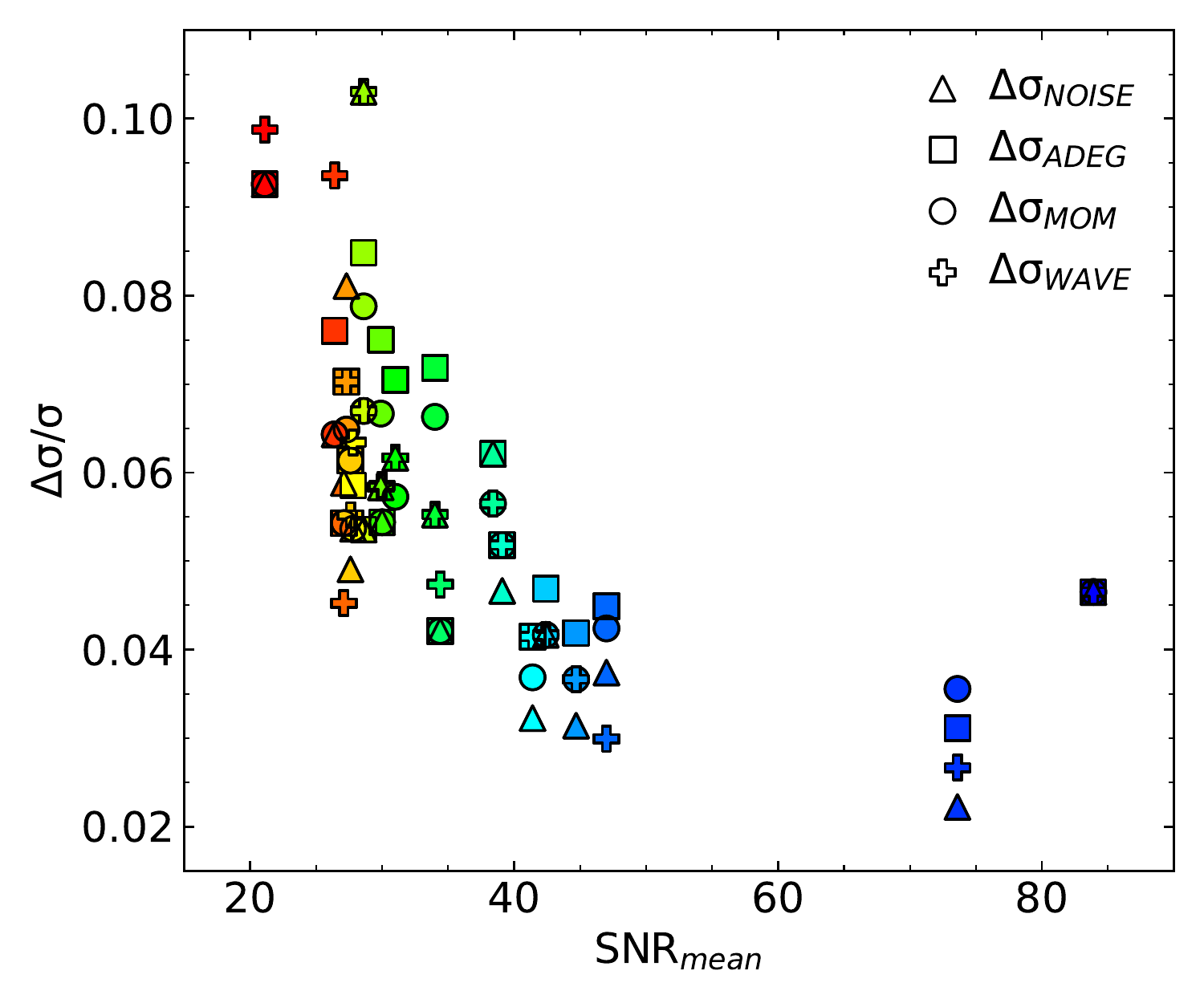}
    \caption{Relative variation of $\sigma$ against the mean S/N of the spectra. The four different symbols show the values relative to the four different bootstrap routines described in Sec.~\ref{sec:bootstrap}, and the points are colour-coded by the S/N of the spectra to which they refer. The uncertainties are about 10\% for the lowest S/N spectra (red) and $\sim4$\% for the highest S/N spectra (blue). The colour gradient of the data points allows distinguishing the different systems.}
    \label{fig:snr-delta}
\end{figure}

All these tests allowed us to control the uncertainties introduced by the choice of the fit parameters during the fit. The $\Delta\sigma$ that are listed in Table~\ref{tab:bootstrap_data} are always $<20$ \kms. Each $\Delta$ column in the table was obtained by also including the previous randomisation(s). Thus, the last column refers to the case where noise, ADEGREE, and MOMENTS were all randomised. 
We note that randomizing on more parameters at the same time does not necessarily correspond to an increase in the associated uncertainty. The four $\Delta\sigma$ columns of Table~\ref{tab:bootstrap_data} all list comparable values, demonstrating that randomizing on all parameters gives a good indication of the total uncertainty associated with all the possible parameters of the fit. Finally, a clear dependence on the S/N of the spectra is found, as shown in Fig.~\ref{fig:snr-delta}. The uncertainties are $\sim10$ (4) \% for the lowest (highest) S/N spectra.  

\begin{table*}
\centering
\begin{tabular}{l|ccccccc}
\hline\hline
  \multicolumn{1}{c|}{ID} &
  \multicolumn{1}{c}{$\sigma$ (\kms)} &
  \multicolumn{1}{c}{$\sigma$ (\kms)} &
  \multicolumn{1}{c}{$\sigma$ (\kms)} &
  \multicolumn{1}{c}{$\sigma$ (\kms)} &
  \multicolumn{1}{c}{$\Delta\sigma_{COMB - UVB}$ } &
  \multicolumn{1}{c}{$\Delta\sigma_{COMB - VIS}$} &
  \multicolumn{1}{c}{$\Delta\sigma_{COMB - NIR}$} \\

  \multicolumn{1}{c|}{KIDS} &
  \multicolumn{1}{c}{UVB}   &
  \multicolumn{1}{c}{VIS}    &
  \multicolumn{1}{c}{NIR}    &
  \multicolumn{1}{c}{COMB}    &
  \multicolumn{1}{c}{(\kms)}    &
  \multicolumn{1}{c}{(\kms)}    &
  \multicolumn{1}{c}{(\kms)}    \\
\hline
J1142+0012 & 112$\pm$7 & 147$\pm$5 & 173$\pm$12 & 129$\pm$3 & 17$\pm$10 & -17$\pm$9 & -44$\pm$16\\ 
J1026+0033 & 208$\pm$4 & 229$\pm$6 & 221$\pm$13 & 225$\pm$3 & 16$\pm$7 & -4$\pm$9 & 4$\pm$16\\ 
J0909+0147 & 365$\pm$8 & 429$\pm$11 & 449$\pm$27 & 401$\pm$7 & 36$\pm$15 & -27$\pm$18 & -47$\pm$33\\ 
J1228-0153 & 177$\pm$5 & 188$\pm$6 & 207$\pm$7 & 191$\pm$3 & 15$\pm$8 & 3$\pm$9 & -16$\pm$10\\ 
J1128-0153 & 185$\pm$5 & 206$\pm$6 & 166$\pm$9 & 192$\pm$4 & 7$\pm$10 & -13$\pm$11 & 26$\pm$14\\ 
J1411+0233 & 197$\pm$5 & 211$\pm$6 & 200$\pm$9 & 217$\pm$3 & 20$\pm$9 & 6$\pm$9 & 17$\pm$12\\ 
J1436+0007 & 188$\pm$5 & 193$\pm$6 & 191$\pm$16 & 193$\pm$4 & 5$\pm$10 & 1$\pm$10 & 2$\pm$21\\ 
J1156-0023 & 166$\pm$6 & 182$\pm$5 & 239$\pm$17 & 177$\pm$4 & 11$\pm$10 & -5$\pm$9 & -62$\pm$21\\ 
J0920+0126 & 168$\pm$7 & 166$\pm$5 & 209$\pm$11 & 190$\pm$6 & 22$\pm$13 & 24$\pm$11 & -19$\pm$17\\ 
J1114+0039 & 154$\pm$7 & 191$\pm$8 & 215$\pm$13 & 181$\pm$8 & 26$\pm$15 & -10$\pm$15 & -35$\pm$21\\ 
J2204-3112 & 212$\pm$8 & 253$\pm$9 & 153$\pm$12 & 227$\pm$7 & 14$\pm$16 & -27$\pm$17 & 74$\pm$20\\ 
J0917-0123 & 225$\pm$11 & 237$\pm$7 & 192$\pm$19 & 239$\pm$6 & 14$\pm$16 & 2$\pm$13 & 47$\pm$24\\ 
J1040+0056 & 227$\pm$12 & 226$\pm$9 & 155$\pm$13 & 240$\pm$8 & 13$\pm$20 & 14$\pm$16 & 85$\pm$20\\ 
J1202+0251 & 169$\pm$9 & 175$\pm$9 & 176$\pm$10 & 165$\pm$6 & -4$\pm$15 & -10$\pm$15 & -11$\pm$16\\ 
J0844+0148 & 195$\pm$9 & 237$\pm$9 & 194$\pm$9 & 224$\pm$6 & 29$\pm$15 & -13$\pm$15 & 30$\pm$15\\ 
J0904-0018 & 206$\pm$10 & 192$\pm$9 & 200$\pm$15 & 205$\pm$6 & -1$\pm$16 & 13$\pm$15 & 4$\pm$21\\ 
J1154-0016 & 136$\pm$7 & 194$\pm$8 & 94$\pm$9 & 163$\pm$4 & 26$\pm$12 & -31$\pm$12 & 69$\pm$14\\ 
J2257-3306 & 170$\pm$6 & 186$\pm$9 & 200$\pm$18 & 185$\pm$6 & 16$\pm$13 & -1$\pm$15 & -15$\pm$25\\ 
J2202-3101 & 187$\pm$10 & 227$\pm$7 & 238$\pm$15 & 221$\pm$5 & 33$\pm$16 & -7$\pm$13 & -17$\pm$20\\ 
J1218+0232 & 168$\pm$8 & 191$\pm$9 & 170$\pm$9 & 171$\pm$6 & 3$\pm$14 & -20$\pm$15 & 1$\pm$15\\ 
J2356-3332 & 148$\pm$9 & 181$\pm$10 & 151$\pm$10 & 162$\pm$6 & 14$\pm$15 & -19$\pm$16 & 12$\pm$16\\ 
\hline\hline
\end{tabular}
\caption{Results of the tests on the input spectra and their resolution. Systems are ordered from the highest to the lowest S/N. We report from left to right the ID, the velocity dispersion values computed from the single-arm spectra at their original resolution and that measured from the combined and smoothed ones (COMB), and the differences in $\sigma$ between these three spectra ($\Delta\sigma$). We note that the uncertainties quoted in the table are just the formal errors produced by  \ppxf. }
\label{tab:kinematic_data}
\end{table*}

\begin{figure*}
\includegraphics[width=18.5cm]{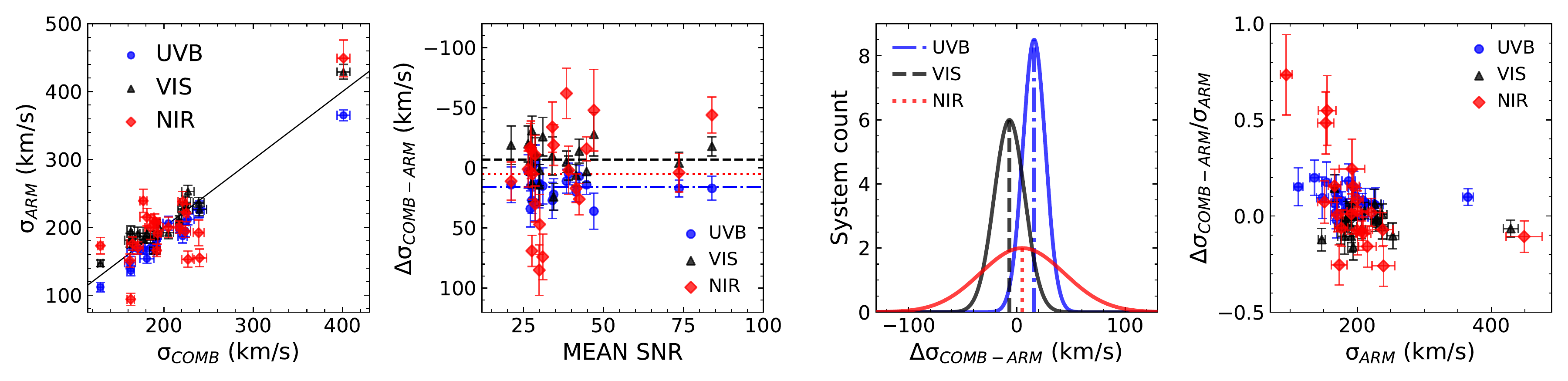}
\caption{Effect of changing the resolution on the $\sigma$ measurement. \textit{First panel (from the left):} Comparison between the velocity dispersion values obtained from the combined spectrum and those obtained from the three single arms at their original resolution. The solid black line shows the one-to-one identity relation. \textit{Second panel:} Difference between the velocity dispersion computed from the combined and smooth spectra and those measured from the single-arm spectra as a function of the mean S/N of each system. The horizontal lines represent the mean offset found for the three arms and correspond to the vertical lines in the following panel (we use a different line style for each different arm). The \textit{y}-axis is flipped to better compare it with the \textit{x}-axis in the third panel. No clear correlation with the S/N is found. \textit{Third panel:}  Distribution of the $\sigma$ differences between the arms, drawn from the histograms, assuming a Gaussian profile. A small offset for the UVB (VIS) is visible, which systematically underestimates (overestimates) the $\sigma$ by 14 \kms\ ($-8$ \kms) on average. For the NIR, the distribution peaks at around 3 \kms,\ but a much larger scatter is found. \textit{Fourth panel:} Relative shift in $\sigma$ against the $\sigma$ measured from the single-arm spectra.}
\label{fig:compar_arms_delta}
\end{figure*}

\subsection{Changing the input spectra and their resolution}
\label{sec:changing_input}
After assessing how the parameters and set-ups of the \ppxf\, code influence the velocity dispersion measurements, we performed one more test by changing the input spectra, both in terms of the considered wavelength band and in terms of spectral resolution and binning. In particular, we ran the code independently on four different spectra for each galaxy: the UVB, the VIS, and the NIR at their original resolution ($R_{\text{UVB}} = 3200$, $R_{\text{VIS}} = 5000$, and $R_{\text{NIR}} = 4300$), and on the combined and smoothed spectrum described in Sec.\ref{sec:smoothing}, up to $\lambda=10000$\AA. 
For this final test, we did not run the bootstrap analysis, did not mask single lines, nor changed the limits of wavelength range on which the fit is performed, as we already know the contribution on the uncertainties that this has. We also fixed ADEGREE to the fiducial value of each system and MOMENTS = 4, given the results of the tests performed so far.  

Overall, the spectral resolution plays a small role, leading to $\sigma$ values consistent within the (statistical) uncertainties. 
This is expected, given the relatively high stellar velocity dispersion values covered by our sample (130-400 \kms), corresponding to a resolution that is well below the resolution of the XSH data and of the MILES models.  
The effect of changing the resolution of the input spectra on the $\sigma$ is shown in the first panel in Fig.~\ref{fig:compar_arms_delta}, where we plot the $\sigma$ values measured from the combined spectrum (between 3000 and 10000 \AA)  on the x-axis and the values obtained from the single arms  on the y-axis. A fair agreement is found between the measurements, especially for UVB and VIS. The measurements in NIR show a much larger scatter, as shown in the second panel of this figure. Here we plot the difference between the $\sigma$ measured from the combined spectrum and that measured from the single arms versus the mean S/N of the corresponding spectrum. Independently of the S/N, the difference in the NIR is larger ($>50$ \kms\ in many cases).  
The third panel of Fig.~\ref{fig:compar_arms_delta} shows the histograms of the $\Delta\sigma$ (Table~\ref{tab:kinematic_data}), from which a much larger scatter is visible from the NIR spectra than from UVB and VIS. This is due to the several spikes, bad pixels, and residuals in this arm. A small systematic difference between the values computed from the UVB ($\sim14$ \kms) and the VIS ($\sim -8$ \kms) is found as well. 
Finally, the rightmost panel demonstrates that the $\sigma$ shift is not a relative effect (e.g. 5\% of the $\sigma$ value), but rather an apparently absolute shift of  20\% in UVB and VIS at most, but up to 60\% in the NIR (for systems  with a lower velocity dispersion and relatively low S/N spectra).

The values of the stellar velocity dispersion measured from the single arms, those measured from the combined and smoothed spectra, and the differences between them are all listed in Table~\ref{tab:kinematic_data}. The errors quoted on the $\Delta\sigma$ are the sum of the errors of the two terms. All the quantities in the table are rounded to the integer. 

The main conclusion that can be drawn from this test is that when different input spectra are used that cover different wavelength ranges, the inferred stellar velocity dispersion values can change systematically and by up to 20\% between UVB and VIS and up to 60\% when NIR is considered as well (however, this happens only for the lower S/N spectra).
We speculate that the different (generally higher) values inferred from blue to red might have a physical origin, especially for non-relics, which might include multiple populations, possibly with different kinematic properties within the galaxy. Excluding the region below $4000$\AA\ might exclude the contribution of relatively younger stars, while focussing on $\lambda>6000$\AA\ might mean that only the oldest and reddest stars are considered. 
Finally, in our case, the spectral region at $\lambda>10000$\AA\ becomes very noisy and does not always allow us to obtain a very precise constraint on the stellar velocity dispersion. 

\begin{figure*}
\centering
\includegraphics[width=\textwidth]{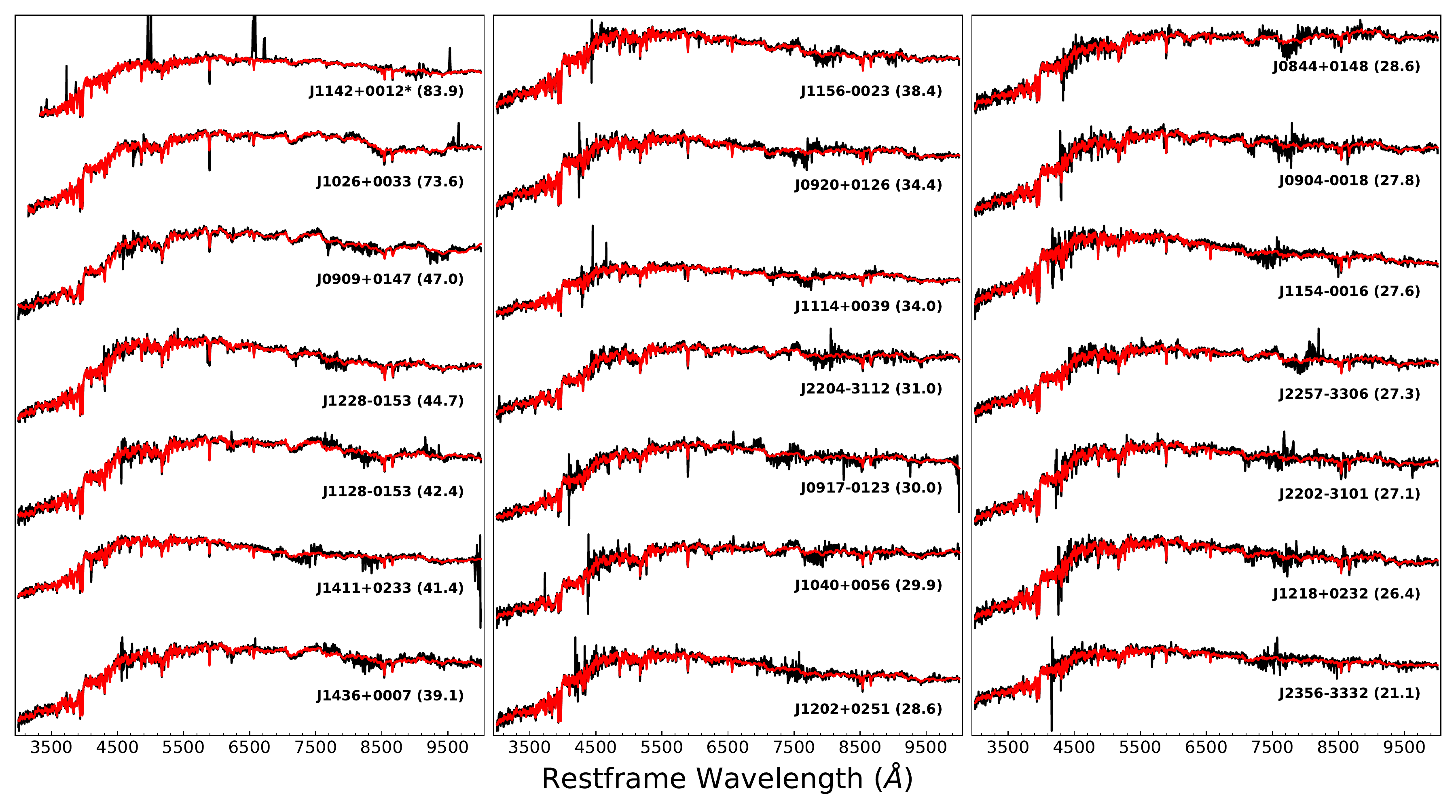}
\caption{Fiducial best fit (red) over-plotted on the combined and smoothed galaxy spectra (red) for all the systems in DR2 ordered from high to low S/N  (left to right and top to bottom). Noisier regions around 4300-4500 \AA\ and around 7500-8000\ \AA\ show the wavelength at which the different arms were joined.\\
\textsuperscript{*}For clarity of the stellar continuum best fit, the peaks of the strong emission lines in the spectrum of J1142+0012 are cut in the plot.}
\label{fig:all_spectra}
\end{figure*}

Hence, in conclusion, the $\sigma$ values computed from the wavelength range [3000-10000] \AA\ probably provide the most robust estimate for XSH spectra with a medium or high S/N. We therefore quote them as our fiducial  choice.
Figure~\ref{fig:all_spectra} shows the fiducial fit for all the systems, grouped and ordered according to the S/N  of the corresponding spectrum. 

\section{Comparison with literature measurements}
\label{sec:meas_comparison}
\begin{figure}
\includegraphics[width=9cm]{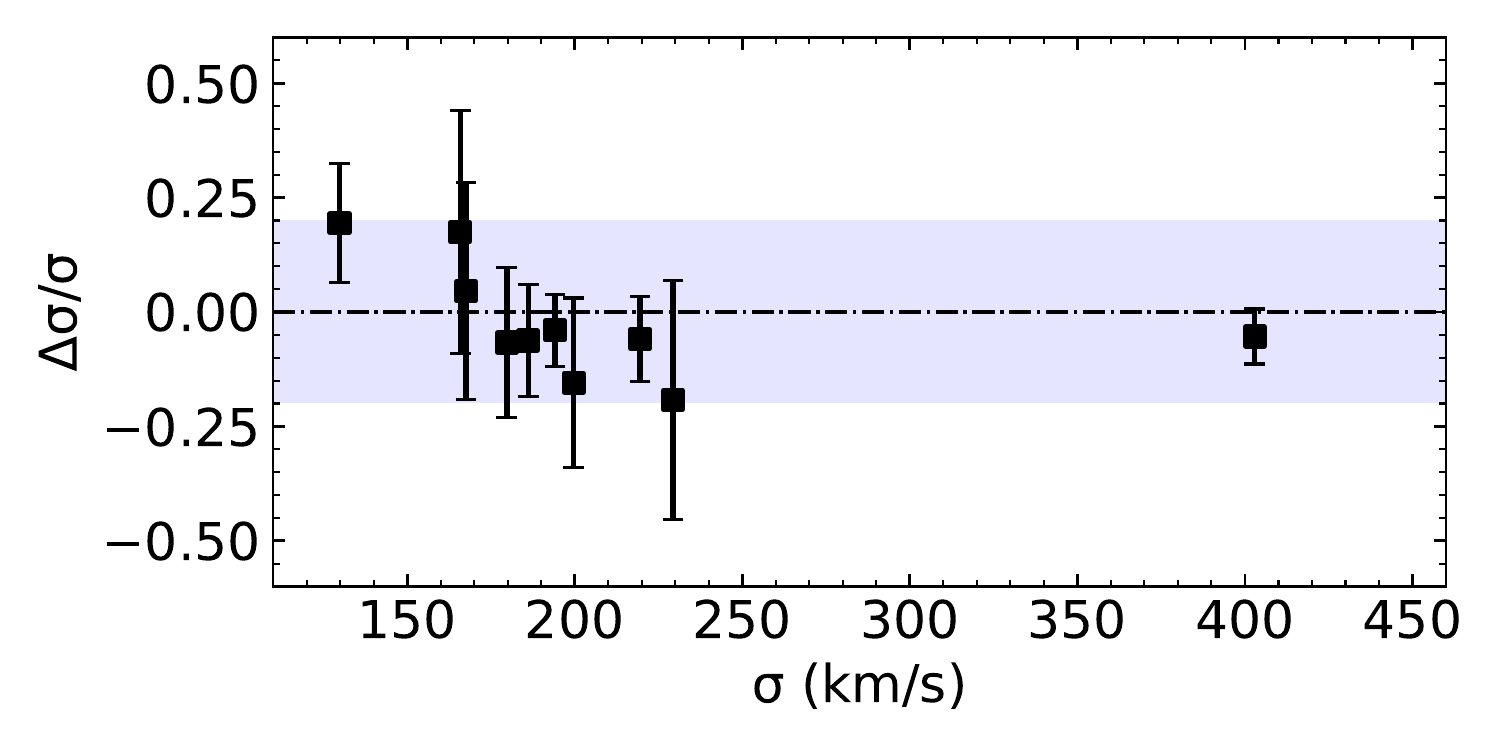}
\caption{Relative difference between the velocity dispersion measured from XSH and that measured from GAMA plotted against the $\sigma$ measured from XSH. The fit was made on the ten objects in common within the same wavelength range and with the same fit settings. The disagreement is about 20\% (shaded region) or more for galaxies with $\sigma<$170 \kms, and it is much smaller for galaxies with a higher velocity dispersion.}
\label{fig:compar_gama}
\end{figure}

Ten of the 21 systems analysed in this \INSPIRE\ DR2 have also been targeted by the GAMA Survey. Generally, the GAMA spectra have a lower or comparable S/N, but much lower (2x) spectral resolution (R$\sim1300$) and thus a larger instrumental dispersion. 

We ran \ppxf\, on these ten GAMA spectra and on the ten \INSPIRE\ spectra by adopting the same fit settings (i.e. the same wavelength range, MOMENTS=4, and the optimal ADEGREE choice for each galaxy), after correcting for the flux shift between the blue and red GAMA arms that is due to a bad splicing. 
The fitted wavelength range for both data sets is $\sim 3500-7000 \AA$. 

We note that the two set of spectra are both fully seeing dominated, and hence the inferred velocity dispersion values should be considered as lower limits (for more details, see Appendix A of the \citetalias{Spiniello+21}). For the \INSPIRE\ case, the R50 apertures are about $\sim 0.5-0.6\arcsec$, and the GAMA spectra are extracted from a circular aperture of radius=$1\arcsec$. 
We decided not to apply any aperture-corrections to the $\sigma$ because the equations derived in the literature (e.g. Eq.~1 of \citealt{Cappellari+06}) for normal-sized galaxies probably cannot be applied to ultra-compact galaxies. However, we note that bringing the measurements to \Reff\ will account for a variation by 5\% of the $\sigma$ at most.

The velocity dispersion values obtained from the XSH and GAMA spectra are compared in Fig.~\ref{fig:compar_gama}. We plot the quantity $\Delta\sigma/\sigma_{\text{XSH}}\equiv(\sigma_{\text{GAMA}}-\sigma_{\text{XSH}})/\sigma_{\text{XSH}}$, representing the relative disagreement between the two values, which are listed in Table~\ref{tab:gama}, along with the uncertainties given by \ppxf\ (only random errors).  
The agreement is fairly good. The velocity dispersion of only one system (J1142+0012) is different by more than 1$\sigma$ error. The scatter is non-negligible for this system, which has the lowest velocity dispersion ($\sigma<150$ \kms) and also the highest S/N. 
Table~\ref{tab:gama} is indeed ordered from the highest to the lowest S/N  spectra, highlighting that $\Delta\sigma$ does not depend on the S/N  of \INSPIRE\ spectra. 

Except for three systems with the lowest velocity dispersion, all the $\Delta\sigma/\sigma$ are negatives. This is in line with the fact that the XSH spectra are integrated over a slightly smaller aperture than the GAMA spectra.

\begin{table}
\centering
\begin{tabular}{l|ccc}
\hline\hline
  \multicolumn{1}{c|}{ID} &
  \multicolumn{1}{c}{$\sigma$ (\kms)} &
  \multicolumn{1}{c}{$\sigma$ (\kms)} &
  \multicolumn{1}{c}{$\Delta\sigma_{XSH - GAMA}$} \\
  \multicolumn{1}{c|}{KIDS} &
  \multicolumn{1}{c}{XSH} &
  \multicolumn{1}{c}{GAMA} &
  \multicolumn{1}{c}{(\kms)} \\
\hline
J1142+0012 & 130$\pm$4 & 155$\pm$12 & 25$\pm$16\\ 
J0909+0147 & 403$\pm$7 & 381$\pm$17 & -22$\pm$24\\ 
J1411+0233 & 219$\pm$3 & 207$\pm$17 & -13$\pm$20\\ 
J1436+0007 & 194$\pm$4 & 186$\pm$11 & -8$\pm$15\\ 
J1156-0023 & 168$\pm$5 & 175$\pm$35 & 8$\pm$40\\ 
J0920+0126 & 186$\pm$5 & 175$\pm$18 & -12$\pm$22\\ 
J0844+0148 & 229$\pm$9 & 185$\pm$49 & -44$\pm$58\\ 
J0904-0018 & 200$\pm$7 & 169$\pm$29 & -31$\pm$36\\ 
J1154-0016 & 166$\pm$5 & 195$\pm$38 & 29$\pm$43\\ 
J2257-3306 & 180$\pm$6 & 168$\pm$23 & -12$\pm$29\\ 
 \hline\hline
\end{tabular}
\caption{Comparison with GAMA. Systems are ordered from highest to lowest S/N . The last column shows the difference between the measurements obtained from the two spectra, using a similar wavelength range and the same ADEGREE and MOMENT parameters. The uncertainty on the $\Delta\sigma$ is assumed to be the sum of the two single uncertainties on the $\sigma$. }
\label{tab:gama}
\end{table}

\begin{figure}
\includegraphics[width=9cm]{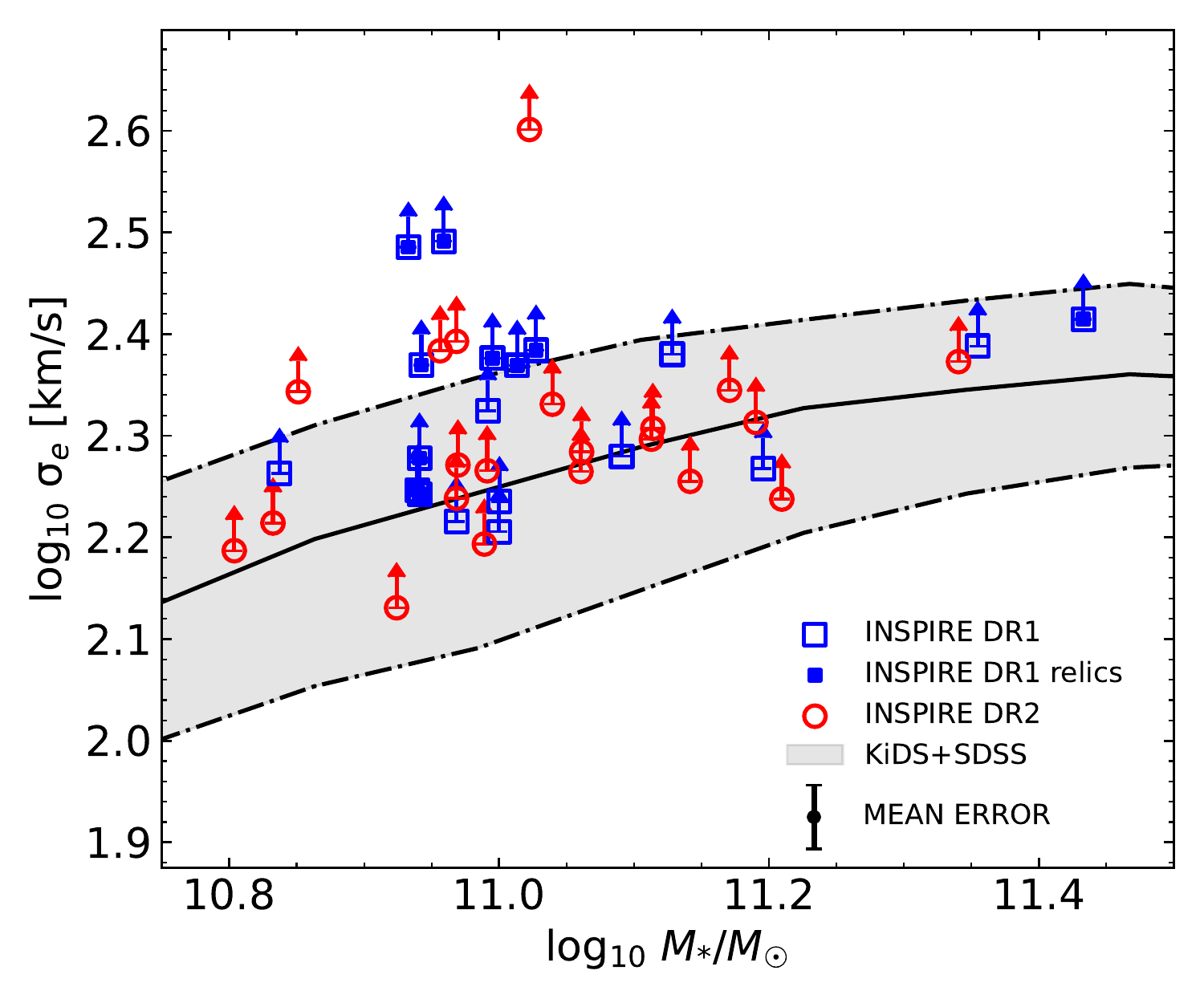}
\caption{ M$_{\star}$--$\sigma$ relation for the INSPIRE DR1+DR2 galaxies compared to normal-sized ETGs. Empty blue squares represent the UCMGs in DR1, and filled blue squares indicate confirmed relics. Empty red circles show the galaxies in DR2. The black line draws the median of the stellar mass-velocity dispersion relation for normal galaxies in KiDS with SDSS DR16 spectroscopy, while the shaded region highlights the 16th--84th percentile interval. Because the \INSPIRE\ spectra are seeing dominated, the estimates of the velocity dispersion values must be considered as lower limits. The black error bar in the legend shows the mean error we retrieve from the bootstrap, and the arrows associated with the single \INSPIRE\ objects show the estimated strength of a 7\% systematic correction that takes the seeing effect into account (see Appendix A of the \citetalias{Spiniello+21}).}
\label{fig:sigma_mass}
\end{figure}

\section{Stellar mass--velocity dispersion relation}
\label{sec:mass-sigma}
In \citetalias{Spiniello+21}, we found a quantitative difference between relics and non-relics in the stellar mass--stellar velocity dispersion space. In particular,  $\sigma$ 
at fixed stellar mass is higher for relics and especially extreme relics than for non-relics and normal-sized passive galaxies. 

We can now reproduce the M$_{\star}$--$\sigma$ plot taking advantage of the increased number statistics that we achieved with DR1+DR2. This plot can potentially have a strong predictive power if we identify that only relics (and not all UCMGs in general) are outliers in this relation, having higher velocity dispersion values. 

The M$_{\star}$--$\sigma$ relation is plotted in Fig.~\ref{fig:sigma_mass} for the 40 \INSPIRE\ objects we analysed so far. 
We recall that the velocity dispersion values measured from the R50 spectra are underestimated because of the seeing effect, and thus must be interpreted as a lower limit to the $\sigma$, as indicated by the arrows. Uncertainties are not drawn for each single system for clarity of the plot, but a mean error, derived from the bootstrapping procedure, is shown in the top left corner of the plot.
Blue points are DR1 galaxies, and red points are the new objects that we added with this paper. The solid black line shows the M$_{\star}$--$\sigma$ relation for a sample of normal-sized ETGs from the KIDS survey for which SDSS DR16 spectroscopy is available. In this case, the velocity dispersion was corrected for \Reff. The shaded region represents the 16th--84th percentile confidence interval. The majority of DR1 confirmed relics (filled squares) generally lie above the median trend for normal galaxies in KiDS, and only DR1 confirmed relics systematically also lie above the 84th percentile. 
We note, nevertheless, that two (non-extreme) relics confirmed in DR1 do not have $\sigma$ that exceeds expectations significantly. 

At least four red points (DR2 UCMGs) appear to have a higher velocity dispersion than normal-size galaxies of similar stellar mass. Hence, we identify them as the most reliable relic candidates from the DR2 sample.  
We therefore conclude that we currently cannot confirm the (tentative) result found in DR1, and that a careful
and detailed stellar population analysis remains the only and best way to confirm the relic nature of ultra-compact massive galaxies. This will be performed on the 21 new objects in a forthcoming paper of the \INSPIRE\ series (Spiniello et al., in prep.).
With a larger number statistic of confirmed relics, we will be able to reassess the situation and finally conclude whether all relics are outliers in the M$_{\star}$--$\sigma$ relation.

\section{Summary and conclusions}
\label{sec:conclusions}
This paper accompanies the \INSPIRE\, second data release (DR2), which is also released as ESO Phase 3 collection (see Sec.~2).
We have reduced and analysed the X-Shooter spectra of an additional 21 UCMGs, selected from the KiDS Survey (\citetalias{Tortora+18_UCMGs, Scognamiglio20}) or from the GAMA Survey, to be good relic candidates. These new 21 spectra are added to those already released in \citet{Spiniello+21} as part of the \INSPIRE\ DR1, and bring the total number of objects analysed so far to 40. 

After reducing the data with the standard ESO XSH pipeline, up to the production of a 2D spectrum, we have extracted the 1D integrated spectra in each arm, from an aperture containing $\sim50$\% of the total light (R50). We note, however, that this does not exactly correspond to extracting spectra at \Reff\ because the light contained in the R50 aperture is a mixture from inside and outside the real \Reff\ being the data seeing limited. In DR1, we showed that the kinematics does not depend on the extraction method.
We then corrected the VIS and NIR for telluric emission using the {\tt molecfit} code and finally obtained a combined UVB+VIS+NIR spectrum for each galaxy after smoothing the spectra to the common resolution of 2.51\AA\, in FWHM. This is the same resolution as for the model spectrum templates.

For each object, we release three final 1D spectra, one for each arm, at the original instrumental resolution (R$=3200$, R$=5000$, and  R$=4300$ in UVB, VIS, and NIR, respectively).  The fluxes are given in units of erg cm$^{-2}$ s$^{-1}$ \AA$^{-1}$, and the wavelength is always measured in air. The combined and smoothed spectrum is also released as an additional ancillary product, together with the original spectra. 
Finally, we release the NIR spectra of the 19 systems presented in DR1, for which UVB and VIS are already publicly available. In this case, we provide both the R50 and the OptExt version for completeness (see DR1 for more information). We add the combined and smoothed spectra as ancillary files for these DR1 objects as well.

The main result of this paper is an in-depth kinematical analysis that demonstrates the validity and robustness of the stellar velocity dispersion measurements and the systematics associated with the derivation of kinematic features with full spectral fitting approaches. We presented the integrated $\sigma$ values obtained from the spectra of the 21 UCMGs and carried out a very detailed quantitative analysis of the statistical and systematic uncertainties on these values due to the different assumptions, parameters, and set-ups of the \ppxf\, code, which was used for the full spectral fitting. 
For the purpose, we also set up a bootstrap analysis, which allowed us to take the statistical uncertainties from the observations and the uncertainties introduced by the specific parameter set-up into account. 

The conclusion of this analysis is that because of the fairly high S/N  of the spectra ($20<S/N _{\text{MEAN}}<85$), 
the stellar velocity dispersion values are robust and generally precise to the 5\% level. However, the wavelength range used in the fit plays a non-negligible role in inferring the $\sigma$ values: it shifts the inferred values also by $\sim30$\% in the worst cases. This mainly depends on the S/N of the input spectrum. 

In detail, we found that the degree of the additive Legendre polynomial that we used to correct for the continuum shape affects the inferred values of $\sigma$ only slightly, unless too low ($<5$) or too high ($>25$) values are used. The number of moments of the Gauss-Hermite parameterisation of the LOSVD used in the fit also plays a negligible role on the stellar velocity dispersion values.
We found a mean systematic shift of $\sim 14$ ($\sim -8$) \kms\ between the stellar velocity dispersion values extracted from the UVB (VIS) spectra and those measured from the combined and smoothed spectra. This shift does not depend strongly on the S/N. The values computed from the NIR are overall consistent, but show a much larger scatter (and hence a larger formal error) because of the noisier nature of the spectra at these redder wavelengths (spikes and sky and telluric residuals) and the absence of strong and narrow absorption lines. The uncertainties computed from the NIR appear to be larger for systems with lower $\sigma$. On the other hand, modifying the wavelength range on which the fit is performed plays the largest role, changing the inferred $\sigma$ by more than the statistical uncertainties returned by the fit and by up to 20\% in the worst cases. The velocity dispersion values are systematically higher when a redder wavelength range is used, within 3000-10000\AA. This mostly happens for lower S/N  spectra, as shown in Fig.~\ref{fig:fitrange_test}, where the velocity dispersion measurements are significantly larger when the blue end of the wavelength is excluded.

We finally analysed the GAMA spectra of ten galaxies in common with the \INSPIRE\ DR2 sample. We repeated the \ppxf\ run on the GAMA and XSH spectra using an identical configuration for the fit. The velocity dispersion agreed fairly well with the dispersion measured from the XSH spectra in a similar wavelength region. Only for systems with the lowest velocity dispersion were the values inferred from GAMA spectra (with a much lower spectral resolution) overestimated by $\sim25\%$.

In conclusion, our work has shown that assessments of stellar kinematics and measurements of the stellar velocity dispersion are mainly robust to different assumptions on the fitting parameters. They are more sensitive to changes in wavelength coverage than previously thought, however. 

In the next paper of the series (Spiniello et al., in prep.), we will focus on the stellar population properties of these 21 UCMGs. Constraining age, metallicity, and [Mg/Fe], we will be able to confirm a fraction of them as relics, further increasing the number of spectroscopically confirmed relics in the low-$z$ Universe. We will therefore test whether the velocity dispersion can be used as a selection criterion to select the most reliable relic candidates among UCMGs. 

Finally, the third and final \INSPIRE\ data release is foreseen after completion of all the observations. Twelve additional UCMGs will be targeted, whose kinematics and stellar populations will be studied. 

\section*{Acknowledgements}
GD acknowledges support by ANID, BASAL, FB210003. CS is supported by an `Hintze Fellowship' at the Oxford Centre for Astrophysical Surveys, which is funded through generous support from the Hintze Family Charitable Foundation.  
CS, CT, FLB, AG, SZ, and PS acknowledge funding from the INAF PRIN-INAF 2019 program 1.05.01.85.11. AFM has received financial support through the Postdoctoral Junior Leader Fellowship Programme from `La Caixa' Banking Foundation (LCF/BQ/LI18/11630007) and from the Severo Ochoa Excellence scheme of the MCIU (CEX2019-000920-S). 
DS is a member of the International Max Planck Research School (IMPRS) for Astronomy and Astrophysics at the Universities of Bonn and Cologne. 

The Geryon cluster at the Centro de Astro--Ingenieria UC was extensively used for the calculations performed in this paper. BASAL CATA PFB--06 and FB210003, the Anillo ACT-86, FONDEQUIP AIC--57, and QUIMAL 130008 provided funding for several improvements to the Geryon cluster.

GAMA is a joint European-Australasian project based around a spectroscopic campaign using the Anglo-Australian Telescope. The GAMA input catalogue is based on data taken from the Sloan Digital Sky Survey and the UKIRT Infrared Deep Sky Survey. Complementary imaging of the GAMA regions is being obtained by a number of independent survey programmes including GALEX MIS, VST KiDS, VISTA VIKING, WISE, Herschel-ATLAS, GMRT and ASKAP providing UV to radio coverage. GAMA is funded by the STFC (UK), the ARC (Australia), the AAO, and the participating institutions. The GAMA website is http://www.gama-survey.org/.

Based on observations made with ESO Telescopes at the La Silla Paranal Observatory under programme ID 179.A-2004 and ID 177.A-3016.

The authors wish to thank the ESO Archive Science Group for the great support with the Data Release. 

We finally thank the anonymous referee for the helpful comments and constructive remarks on this manuscript.

%%%%%%%%%%%%%%%%%%%%%%%%%%%%%%%%%%%%%%%%%%%%%%%%%%



\bibliographystyle{aa}
\bibliography{aa45542-22.bib} 

\end{document}